\documentclass[aps,twocolumn,amsmath,amssymb,nofootinbib,preprintnumbers,showpacs]{revtex4}
\usepackage{epsfig}
\usepackage{bbm}
\usepackage{mathrsfs,graphicx,bm,amsmath}

\newcommand{\Tr}{\mathrm{Tr}}

\newcommand{\hpn}{\tilde{h}_\phi}
\newcommand{\Tn}{\tilde{T}}

\newcommand{\hF}{^{(F)}}
\newcommand{\hB}{^{(B)}}
\newcommand{\lit}{^6\mathrm{Li}}
\newcommand{\kal}{^{40}\mathrm{K}}
\newcommand{\epm}{\epsilon_{\text{M}}}

\newcommand{\tih}{\tilde{h}}
\newcommand{\lan}{\tilde{\lambda}_\phi}
\newcommand{\tila}{\tilde{\lambda}}

\newcommand{\hph}{\hat{h}_\phi}

\newcommand{\nuh}{\hat{\nu}}
\newcommand{\lpsih}{\hat{\lambda}_\psi}
\newcommand{\kh}{\hat{k}}
\newcommand{\sigmah}{\hat{\sigma}}
\newcommand{\ba}{\begin{eqnarray}}
\newcommand{\ea}{\end{eqnarray}}
\newcommand{\be}{\begin{equation}}
\newcommand{\ee}{\end{equation}}

\def\di{\displaystyle}

\def\bg{\begin{eqnarray}\begin{array}{rcl}\displaystyle}
\def\eg{\end{array} &\di    &\di   \end{eqnarray}}
\def\bm#1{\begin{eqnarray}\begin{array}{#1}\di}
\def\bmo#1{\begin{eqnarray*}\begin{array}{#1}\di}
\def\bml#1#2{\begin{eqnarray}\begin{array}{#1}\label{#2}\di}
\def\bgo{\begin{eqnarray*}\begin{array}{rcl}\displaystyle}
\def\ego{\end{array} &\di    &\di \nonumber  \end{eqnarray*}}

\def\btensor#1#2{\renew\left#1\begin{array}{#2}\di}
\def\brtensor#1#2#3{\ren#3\left#1\begin{array}{#2}}
\def\botensor#1#2{\renew\left#1\begin{array}{#2}}
\def\etensor#1{\end{array}\right#1}

\def\eq#1{(\ref{#1})}

\def\Tr{{\rm Tr}}

\def\s0#1#2{\mbox{\small{$ \frac{#1}{#2} $}}}
\def\0#1#2{\frac{#1}{#2}}

\begin{document}

\title{Renormalisation Flow and Universality for Ultracold Fermionic Atoms}

\author{S.~Diehl${}^{a}$}
\author{H.~Gies${}^{b}$}
\author{J.~M.~Pawlowski${}^{b}$}
\author{C.~Wetterich${}^{b}$}

\affiliation{\mbox{\it ${}^a$Institute for Quantum Optics and Quantum
Information of the Austrian Academy of Sciences,}\\
\mbox{\it A-6020 Innsbruck, Austria}\\
\mbox{\it ${}^b$Institut f{\"u}r Theoretische Physik,
Philosophenweg 16, D-69120 Heidelberg, Germany}}


\begin{abstract}
  A functional renormalisation group study for the BEC-BCS crossover
  for ultracold gases of fermionic atoms is presented. We discuss the
  fixed point which is at the origin of universality for broad
  Feshbach resonances.  All macroscopic quantities depend only on one
  relevant parameter, the concentration $ak_{\text{F}}$, besides their
  dependence on the temperature in units of the Fermi energy. In
  particular, we compute the universal ratio between molecular and
  atomic scattering length in vacuum. We also present an estimate to
  which level of accuracy universality holds for gases of Li and K
  atoms.
\end{abstract}


\maketitle

\section{Introduction}
The quantitatively precise understanding of the crossover from a
Bose-Einstein condensate (BEC) to BCS superfluidity \cite{ALeggett80}
in gases of ultracold fermionic atoms \cite{DStoof96,Chen04} is a
theoretical challenge. If it can be met, the comparison with future
experimental precision results could set a new milestone for the
understanding of the transition from known microscopic laws to
macroscopic observations at length scales several orders of magnitude
larger than the characteristic atomic or molecular length scales.
Furthermore, these systems could become a major testing ground for
theoretical methods dealing with the fluctuation problem in complex
many-body systems in a context where no small couplings are available.
The theoretical progress could go far beyond the understanding of
critical exponents, amplitudes and the equation of state near a
second-order phase transition. Then, a whole range in temperature and
coupling constants could become accessible to precise calculations and
experimental tests \cite{Jin04}.

The qualitative features of the BEC-BCS crossover through a Feshbach
resonance can already be well reproduced by extensions of mean field
theory which account for the contribution to the density from di-atom
or molecule collective or bound states \cite{Stoof05}. Furthermore, in
the limit of a narrow Feshbach resonance the crossover problem can be
solved exactly \cite{Diehl:2005an}, with the possibility of a
perturbative expansion for a small Feshbach or Yukawa coupling. A
systematic expansion beyond the case of small Yukawa couplings has
been performed as a $1/N$-expansion \cite{Sachdev06}. The crossover
regime is also accessible to $\epsilon$-expansion techniques
\cite{Nishida:2006br}. At zero temperature \cite{Giorgini04} as well
as at finite temperature \cite{Bulgac:2005pj}, numerical results based
on various Monte-Carlo methods are available at the crossover. A
unified picture of the whole phase diagram has arisen from functional
field-theoretical techniques, in particular from self-consistent or $t$-matrix approaches
\cite{StrinatiPieri04}, Dyson-Schwinger equations
\cite{Diehl:2005an,Diehl:2005ae}, 2PI methods \cite{Rantner}, and the
functional renormalisation group (RG) \cite{Diehl:2007th}.

Recently, it has been advocated \cite{Diehl:2005an} that a large
universality holds for broad Feshbach resonances: all purely
``macroscopic'' quantities can be expressed in terms of only two
parameters, namely the concentration $c=ak_{\text{F}}$ and the
temperature in units of the Fermi energy,
$\hat{T}=T/\epsilon_{\text{F}}$. Here, $k_{\text{F}}$ and
$\epsilon_{\text{F}}$ are defined by the density of atoms,
$n=k^3_{\text{F}}/(3\pi^2),\epsilon_{\text{F}}=k^2_{\text{F}}/(2M)$
and $a$ is the scattering length. Universality means that the
thermodynamic quantities and the correlation functions can be computed
independently of the particular realizations of microscopic physics,
as for example the Feshbach resonances in $\lit$ or $\kal$. A special
point in this large universality region is the location of the
resonance, $c\rightarrow\infty$, where the scaling argument by Ho
\cite{Ho104} applies.

We emphasise that the universal description in terms of two parameters
holds even in situations where the understanding of the microscopic
physics may necessitate several other parameters. It is thus much more
than the simple observation that the appropriate model involves only
two parameters. The fact that two effective parameters are sufficient
is related to the existence of a fixed point in the renormalisation
group flow; for a given $T$, this fixed point has only one relevant
direction, namely the concentration $c$. Then, the question of
accuracy of a two-parameter description is related to the rate how
fast this fixed point is approached by the flow towards the
infrared. It depends on the scales involved and the physical questions
under investigation. As in statistical physics, the deviations from
universality can be related to the scaling dimensions of operators
evaluated at infrared fixed points.  We will address these
questions quantitatively for Feshbach resonances in $\lit$ or $\kal$.

Even for broad Feshbach resonances not all quantities admit a
universal description.  An example is given by the number density of
microscopic molecules which depends on further parameters
\cite{ChenLevin05,Stoof05,Diehl:2005ae}.

The quantitative study of this universality has first been performed
through the solution of Schwinger-Dyson equations, including the
molecule fluctuations \cite{Diehl:2005an,Diehl:2005ae,Diehl:2007tg}.
Already in this work it has been argued that the key for a proper
understanding of universality lies in the renormalisation flow and its
(partial) fixed points. A similar point of view has been advocated by
Nikolic and Sachdev \cite{Sachdev06}. A first functional
renormalisation group study evaluating the renormalisation flow for
the whole phase diagram has been put forward in \cite{Diehl:2007th}.
In the present paper, we perform a systematic study of the
universality aspects of the renormalisation flow for various couplings
thereby extending the results of \cite{Diehl:2007th} concerning
universality. We relate the universality for broad Feshbach resonances
to a non-perturbative fixed point which is infrared stable except for
one relevant parameter corresponding to the concentration $c$.

Our approach is based on approximate solutions of exact functional
renormalisation group equations \cite{CWRG}, for reviews see
\cite{Tetradis,Aoki:2000wm,Pawlowski:2005xe}, and for applications to
non-relavtivistic fermions see \cite{Birse:2004ha}. These are derived
by varying an effective ``infrared'' cutoff associated to a momentum
scale $k$ (in units of $k_{\text{F}}$). The solution to the
fluctuation problem corresponds then to the limit $k\rightarrow 0$. In
this way, a continuous interpolation from the microscopic Hamiltonian
or classical action to the macroscopic observables described by the
effective action is achieved. In this paper, we will use a very simple
cutoff associated to an additional negative chemical potential. This
works well on the BEC side of the crossover and in the vacuum limit of
vanishing density and temperature. Since the crucial characteristics
of the fixed points relevant for universality can be found in the
vacuum limit, this will be sufficient for our purpose. The
investigation of the whole phase diagram in \cite{Diehl:2007th} has
been performed with a different cutoff more amiable to this global
task, see \cite{Litim:2000ci,Pawlowski:2005xe}.

The simple form of the cutoff allows for analytic solutions of the
flow equations; these can be used for a direct check of the more
general arguments, resulting from the investigation of fixed points
and their stability. As a concrete example, we compute the universal
ratio of the molecular scattering length to the atomic scattering
length in vacuum. This ratio is reduced as compared to its mean field
value by the presence of fluctuations of collective bosonic degrees of
freedom. We also present quantitative estimates to what precision
universality is realized for ultracold gases of Li and K.

In Sect.~\ref{sec:func}, we introduce the functional-integral
formulation of our model for the Feshbach resonance. It contains
explicit bosonic fields for di-atom states
\cite{BBKokkelmans,EEGriffin,WWStoofBos}. In the limit of a
divergent Feshbach or Yukawa coupling $\hat{h}_\varphi$, this model is
equivalent to a purely fermionic formulation with a point-like
interaction. In Sect.~\ref{sec:exact}, we briefly recapitulate the
framework of the exact flow equation for the effective
average action, and specify our cutoff and truncations. The initial
values of the flow at the microscopic scale are given in
Sect.~\ref{initial}, while explicit formulae for the flow of couplings
and the effective potential are computed in Sect.~\ref{running} and
\ref{flow}.

In Sect.~\ref{solving}, we specialise to $T=0$ and solve the flow
equations for various couplings explicitly within our
truncation. Sect. \ref{flowandfixed} generalises these results and
presents a general discussion of the fixed points in the flow of the
rescaled dimensionless parameters $\tilde{h}_\varphi^2$,
$\tilde{m}_\varphi^2$ and $\tilde{\lambda}_\psi$, corresponding to the
Yukawa coupling, the energy difference between open- and closed-channel
states and an additional point-like four-fermion vertex which accounts
for a ``background scattering length''. In particular, the fixed point
describing broad Feshbach resonances has only one relevant parameter
$c$, whereas the fixed point characteristic for narrow Feshbach
resonances has two relevant parameters $c$ and
$\tilde{h}_\varphi^2$. Sect. \ref{density} supplements a discussion
of the density which is needed for a practical contact to experiment,
and Sect.~\ref{vacuum} relates the initial values of the flow (the
classical action) to observable quantities, such as the molecular
binding energy or the scattering length depending on a magnetic
field $B$. 

In Sect.~\ref{sec:scat}, we discuss the fixed-point behaviour for an
additional parameter, namely the effective four-boson vertex
$\lambda_\varphi$. This fixed point is responsible for the universal
ratio between the scattering length for molecules and for atoms. We
finally present a general discussion of universality for ultracold
fermionic atoms in Sect.~\ref{universality}, and we estimate the
deviations from the universal broad-resonance limit for Li and K in
Sect.~\ref{deviations}. Sect. \ref{conclusion} contains our
conclusions.

\section{Functional integral}
\label{sec:func}

We start from the partition function in presence of sources ($\int
d\hat x = \int d^3 \hat{\vec{x}} d \hat \tau$)
\begin{eqnarray}\label{YukawaAction}
Z_B[j_\phi]&=&\int {\cal D}\hat\psi {\cal D}\hat{\phi}\exp
\Big\{-S [\hat{\psi},\hat{\phi}]\\\nonumber
&&\quad +\int d\hat x\big[{j}_\phi^*(x)\hat\phi(x) +
  {j}_\phi(x)\hat\phi^*(x)\big]\Big\}, 
\end{eqnarray}
with 
\begin{eqnarray*}
S\hspace{-0.15cm}&=&\hspace{-0.15cm}\int \hspace{-0.12cm} d\hat x
\Big[\hat\psi^\dagger\big(\hat\partial_{\tau}
-\hat\triangle -\hat\sigma\big)\hat\psi+ \frac{(\hat{\lambda}_\psi +
  \delta \hat{\lambda}_\psi)}{2}
(\hat\psi^\dagger\hat\psi)^2\nonumber\\
&&+\hat{\phi}^*\big(\hat\partial_\tau -\frac{\hat\triangle}{2} +
\hat{\nu} +\delta\hat\nu - 2\hat\sigma\big)\hat{\phi} \nonumber\\
&&\hspace{-0.12cm}-(\hat h_\varphi + \delta \hat h_\varphi
)\Big(\hat{\phi}^*\hat\psi_1\hat\psi_2
- \hat{\phi}\hat\psi^*_1\hat\psi^*_2\Big)
\Big].\nonumber
\end{eqnarray*}
Here, we have rescaled the fields and couplings, together with the
space and time coordinates $\hat{\vec{x}} = \hat k \vec x$, $\hat \tau
= (\hat k^2/2M)\tau$, with $M$ being the mass of the atoms. Our units
are $\hbar = c = k_B =1$. We use the Matsubara formalism with
Euclidean time $\tau$ on a torus with a circumference given by the
inverse temperature $T^{-1}$.  The thermodynamic variables are $\hat T
= 2M T /\hat k^2$ and $\hat\sigma = 2M\sigma /\hat k^2$, with $\sigma$
denoting the effective chemical potential.

The model parameters are: 
\begin{itemize}
\item[(1)] the detuning of the magnetic field $B- B_0$ (with $\bar\mu =2
    \mu_B$ for $\lit$ and $\bar\mu = 1.57\mu_B$ for $\kal$)
\begin{eqnarray}
\hat\nu = \frac{2M}{\hat k^2} \bar\mu (B-B_0),
\end{eqnarray}
\item[(2)] the Feshbach or Yukawa coupling $\hat h_\varphi$ which can be
extracted from the properties of the quantum mechanical two-atom
system as the molecular binding energy or the scattering cross
section,  
\item[(3)] a point-like interaction for the fermionic atoms parameterised by
$\hat\lambda_\psi$. The shifts $\delta\nuh$, $\delta
\hat{\lambda}_\psi$ and $\delta\hph$ are counter terms that are
removed by the ultraviolet renormalisation. Details of the rescaling
and the formulation can be found in \cite{Diehl:2005ae,Diehl:2007th}.
\end{itemize}
The scale $\kh$ (which sets the units) is arbitrary. (A typical value
is $\hat k = 1\mathrm{eV}$.)
Under a rescaling of $\kh \to  \hat{k}/\mu$ all quantities scale
according to their canonical dimension, i.e.,
\begin{eqnarray}\label{4}
\hat{x} &=& \hat{k} x \to \hat{x}/\mu, \quad 
\hat{\tau} = (\hat{k}^2/(2M)) \tau \to \hat{\tau}/\mu^2,\nonumber\\
\hat{T} &\to& \mu^2 \hat{T}, \quad
\sigmah \to \mu^2\hat{\sigma},\quad \nuh \to \mu^2\nuh, \quad
\hph\to \mu^{1/2} \hph, \nonumber\\
\hat{\lambda}_\psi &\to& \lpsih /\mu, \quad 
\hat{\psi} \to \mu^{3/2}\hat\psi, \quad
\hat{\phi} \to \mu^{3/2} \hat\phi.
\end{eqnarray}
We observe that the canonical dimension of time is minus two and
therefore the nonrelativistic Lagrangian has dimension five and not
four, as for a relativistic quantum field theory. 
The total atom density $n$ defines the Fermi momentum $k_{\text{F}}$,
\begin{eqnarray}\label{FermiDef}
n = \frac{k_{\text{F}}^3}{3\pi^2}.
\end{eqnarray}
If one associates $\hat{k}$ with $k_{\text{F}}$ all quantities are
expressed in units of the Fermi momentum \cite{Diehl:2005ae}.

From the partition function the effective action $\Gamma$ obtains by
the usual Legendre transform, $\hat\varphi = \langle\hat\phi\rangle$,
\begin{eqnarray}
\Gamma[\hat\varphi ] = -\ln Z[j_\phi[\hat\varphi ]] + \int d\hat x
({j}_\phi^*\hat\varphi + {j}_\phi\hat\varphi^*). 
\end{eqnarray}
The order parameter $\hat{\varphi}_0$ for superfluidity corresponds to
the minimum of $\Gamma$ for $j_\phi =0$ and obeys the field equation
\begin{eqnarray}\label{FieldEq}
\frac{\delta \Gamma}{\delta \hat{\varphi}}\Big|_{\hat{\varphi}_0} = 0.
\end{eqnarray}
The effective action generates the 1PI Green's functions such that the
propagators and transition amplitudes can be directly related to the
functional derivatives of $\Gamma$. The construction above is easily
extended to an effective action that is also a functional of fermionic
fields by introducing the appropriate fermionic sources, see below.

\section{Exact functional flow equation and truncation}
\label{sec:exact}

The variation of $\Gamma$ with the change of an effective infrared
cutoff $k$ is given by an exact renormalisation group equation
\cite{CWRG,Tetradis,Aoki:2000wm,Pawlowski:2005xe}. For the present
theory, this approach has been implemented in \cite{Diehl:2007th}
within a study of the phase diagram; see also \cite{Birse:2004ha}. For RG studies of the 
purely bosonic system, see \cite{Blaizot:2006vr}. 

For the purpose of the present work, we extend the truncation used in
\cite{Diehl:2007th} as well as using a particular version of the flow
equation where the cutoff acts like a shift in the respective chemical
potentials for $\psi$ and $\varphi$, as also done
in \cite{Jungnickel03}. To that end, we introduce an
infrared-regularised partition function $Z_k$ by adding a cutoff term
to the action in Eq.~\eqref{YukawaAction},
\begin{equation}
S[\hat{\psi},\hat{\phi}] \to
S[\hat{\psi},\hat{\phi}] +\Delta S_k [\hat{\psi},\hat{\phi}],
\end{equation}
where the cutoff term $\Delta S_k$ is chosen as
\begin{equation}\label{9A}
\Delta S_k[\hat{\psi},\hat{\phi}]
=\int d\hat{x}(R^{(F)}_k\hat{\psi}^\dagger\hat{\psi}
+R^{(B)}_k\hat{\phi}^*\hat{\phi}), 
\end{equation}
with
\begin{eqnarray}\label{9B}
  R^{(F)}_k&=&Z_\psi(k)k^2~,~R^{(B)}_k=2Z_\varphi(k)k^2,\nonumber\\
  \frac{\partial R^{(F)}_k}{\partial k^2}&=&Q_\psi
  Z_\psi~,~\frac{\partial R^{(B)}_k}{\partial k^2}=2Q_\varphi
  Z_\varphi,\nonumber\\ 
  Q_{\psi,\varphi}&=&1-\eta_{\psi,\varphi}/2~,
  ~\eta_{\psi,\varphi}=-k\frac{\partial}{\partial k}\ln Z_{\psi,\varphi}.
\end{eqnarray}
The $k$-dependent wave function renormalisation $Z_{\psi,\varphi}$
will be determined below, and we note that $k$ is measured in units of
the fixed scale $\hat{k}$. Introducing the effective average action
$\Gamma_k$ in terms of a modified Legendre transform,
\begin{eqnarray}
  \Gamma_k[\hat\psi,\hat\varphi]&=& -\ln Z_k[j_\psi,j_\phi] -\Delta
  S_k[\hat\psi,\hat\varphi] \nonumber\\
  &&+\int d\hat{x} (j_\phi^* \hat\varphi +j_\phi \hat{\varphi}^* 
  + j_\psi^\dagger \hat{\psi}-\hat{\psi}^\dagger j_\psi),
\label{eq:defgamk}
\end{eqnarray}
the exact functional renormalisation group
equation (flow equation) can straightforwardly be derived,
\begin{eqnarray}\label{ExFlowEq}
  \frac{\partial\Gamma_k}{\partial k^2} &=& \int d \hat x \big(Z_\psi
  Q_\psi \langle\hat\psi^\dagger\hat\psi \rangle_c+
  2 Z_\varphi Q_\varphi\langle\hat\phi^*\hat\phi \rangle_c\big)
  \nonumber\\
  &=& \Tr \big\{ -Z_\psi
  Q_\psi\big(\Gamma^{(2)}_k+R_k\big)^{-1}_{\hat\psi^*\hat\psi}
  \nonumber\\
  &&\qquad + 2 Z_\varphi
  Q_\varphi\big(\Gamma^{(2)}_k+R_k\big)^{-1}_{\hat\varphi^*
    \hat\varphi}\Big\}.
\end{eqnarray}
Here, we have expressed the propagators of the fermionic and the
bosonic fields by the corresponding components of the inverse of the
matrix of second functional derivatives of $\Gamma_k$. The trace Tr
contains an integration over $\hat x$ or a corresponding momentum
integration as well as a trace over all internal indices. Both
$\Gamma_k$ and $\Gamma^{(2)}_k$ are functionals of arbitrary fields
$\hat\varphi$ and $\hat\psi$ which are kept fixed for the $k$
derivative in Eq.~(\ref{ExFlowEq}).

The flow equation (\ref{ExFlowEq}) is a nonlinear functional
differential equation and we can only hope to find approximate
solutions by suitable truncations of the most general form of
$\Gamma_k$. In this work, we exploit the ansatz
\begin{eqnarray}\label{Trunc}
  \Gamma_k\hspace{-0.15cm}&=&\hspace{-0.15cm}\int \hspace{-0.12cm} 
  d\hat x \Big[Z_\psi\hat\psi^\dagger\big(\hat\partial_{\tau}
  -A_\psi\hat\triangle -\hat\sigma\big)\hat\psi\nonumber\\
  &&+Z_\varphi\hat{\varphi}^*\big(\hat\partial_\tau - A_\varphi\hat
  \triangle\big)\hat{\varphi} + u(\hat\varphi,\hat{\sigma}) \nonumber\\
  &&\hspace{-0.12cm}-\hat h_\varphi\Big(\hat{\varphi}^*\hat\psi_1
  \hat\psi_2
  - \hat{\varphi}\hat\psi^*_1\hat\psi^*_2\Big) +
  \frac{\hat{\lambda}_\psi}{2}(\hat\psi^\dagger\hat\psi)^2\Big]
  \nonumber\\
  &=& \hspace{-0.15cm}\int \hspace{-0.12cm} d\hat x \Big[\psi^\dagger
  \big(\hat\partial_{\tau}
  -A_\psi\hat\triangle -\hat\sigma\big)\psi\nonumber\\
  &&+ \varphi^*\big(\hat\partial_\tau - A_\varphi\hat\triangle\big)
  \varphi + u(\varphi,\hat{\sigma}) \nonumber\\
  &&\hspace{-0.12cm}- h_\varphi\Big({\varphi}^*\psi_1\psi_2
  - {\varphi}\psi^*_1\psi^*_2\Big) +
  \frac{\lambda_\psi}{2}(\psi^\dagger\psi)^2\Big], 
\end{eqnarray}
In the second equation, we have used renormalised fields
\begin{eqnarray}
  \psi = Z_\psi^{1/2}\hat\psi,\quad  \varphi = Z_\varphi^{1/2}
  \hat\varphi
\end{eqnarray}
and renormalised couplings
\begin{eqnarray}
  h_\varphi = \hat{h}_\varphi Z_\psi^{-1} Z_\varphi^{-1/2},  
  \quad\lambda_\psi = \hat{\lambda}_\psi Z_\psi^{-2}.
\end{eqnarray}
The ansatz \eq{Trunc} generalises that of \cite{Diehl:2007th} by the
four-fermion interaction with coupling $\lambda_\psi$. This term is
particularly important in the limit of broad Feshbach resonances as
will be discussed later.

The global symmetry of $U(1)$ phase rotations implies that the
effective potential $u$ can only depend on $\rho=\varphi^*\varphi$.
If the ground state corresponds to a homogeneous $\varphi_0$ the
pressure and density can be computed from the properties of the
effective potential
\begin{eqnarray}\label{11A}
  \frac{\partial u}{\partial\varphi}{\Big|_{\varphi_0}}&=&0~,~\quad
  u_0(\hat{\sigma})=u(\varphi_0,\hat{\sigma})~,~\nonumber\\
  p&=&-\frac{\hat k^5}{2M}u_0~,\quad ~n=-\hat k^3
  \frac{\partial u_0}{\partial\hat \sigma}.
\end{eqnarray}

At this point, some comments concerning the wave function
renormalisations are in order: 
\begin{itemize}
\item[(i)] Analytic continuation to ``real
time'' and a Fourier transform to (real) frequency space results in
$\partial_\tau \to - \omega$. We define $Z_{\psi,\varphi}$ by the
coefficient of the term linear in $\omega$ in the full inverse
propagator $\hat P_{\psi,\varphi}$ given by the terms quadratic in
$\hat\psi$ or $\hat\varphi$ in $\Gamma_k$. More precisely, we choose
in Fourier space
\begin{eqnarray}\label{PrecZDef}
  Z_\varphi = -\frac{\partial \hat P_\varphi}{\partial \omega}
  \Big|_{\vec q =0, \omega=0}.
\end{eqnarray}
\item[(ii)] With this definition of $Z_\psi$ and $Z_\varphi$, the
renormalised fields $\psi,\varphi$ have a unit residuum for the pole
in the propagator for $\vec q\rightarrow 0$ if the ``on shell value''
of $\omega$ vanishes for $\vec q =0$.  
\item[(iii)] We use the same
$Z_{\psi,\varphi}$ in the definition of the cutoff (\ref{9B}) as in
Eq.~(\ref{Trunc}). 
\item[(iv)] The fields $\hat \psi$ and $\hat \varphi$
describe ``microscopic'' or ``bare'' atoms and molecules while the
renormalised field $\varphi$ describes dressed molecules. The wave
function renormalisation $Z_\varphi$ accounts for a description of
dressed molecules as a mixing of microscopic molecules and di-atom
states \cite{Stoof05,Diehl:2005an,Diehl:2007th}. 
\end{itemize}

\section{Initial conditions}
\label{initial}

Inserting the truncation \eq{Trunc} into the flow equation
\eq{ExFlowEq}, and taking appropriate functional derivatives, leads to
a coupled set of differential equations for the couplings $Z_\psi,
A_\psi, Z_\varphi, A_\varphi, h_\varphi, \lambda_\psi$ as well as the
effective potential $u(\varphi)$.  The task is to follow the flow of
these couplings as the cutoff scale $k$ is changed. The initial values
for large $k$ are taken as
\begin{eqnarray}\label{13}
  Z_\varphi= Z_\psi &=& ~1~,\qquad   A_\varphi = \frac{1}{2},
  \quad A_\psi = 1,\nonumber\\
  u &=& m_\varphi^2\rho,\nonumber\\
  m_\varphi^2 &=& \nu + \delta\nu - 2 \hat\sigma.
\end{eqnarray}
Here $\nu$ is related to the detuning of the magnetic field
\begin{eqnarray}\label{14}
  \nu &=& \frac{2M}{\kh^2} \bar{\mu}(B- B_0) \\\nonumber
\end{eqnarray}
and we have to fix $\delta\nu$ (i.e., the renormalised counterpart of
$\delta \hat{\nu}$ in Eq.~\eqref{YukawaAction}) such that the Feshbach
resonance in vacuum occurs for $\nu=0$. The initial values of
$\lambda_\psi$ and $ h_\varphi$ are chosen such that the molecular
binding energy and scattering of two atoms in vacuum is correctly
described.

A realistic model contains an effective ultraviolet cutoff $\Lambda$
which is given roughly by the inverse of the range of the van der
Waals force. For $k\ll\Lambda/\hat{k}$ we can take the limit
$\Lambda\rightarrow\infty$ since all momentum integrals are
ultraviolet finite. All ultraviolet ``divergencies'' are already
absorbed in the computation of $\Gamma_\Lambda$
\cite{Diehl:2005ae,Diehl:2007th}.  For $\lit$ or $\kal$ and
$\hat{k}=1\mathrm{eV}$ one has $\Lambda/\hat{k}$ of order $10^2$ to $10^3$.

\section{Running couplings and anomalous dimensions}
\label{running}
The flow equations for the various couplings and the densities derived
from \eq{ExFlowEq} have a simple interpretation as renormalisation
group improved one-loop equations \cite{CWRG} with full propagators
and dressed vertices, but are exact. The renormalisation constants
$Z_\varphi, Z_\psi$ are related to the dependence of the
unrenormalised inverse propagators (for $\hat\varphi$ and $\hat\psi$)
on the (Minkowski) frequency, i.e., the coefficients of the terms
linear in $\omega(\hat{=}-\partial/
\partial\tau)$ in $\Gamma^{(2)}$. The second derivative of
Eq.~(\ref{ExFlowEq}) yields exact flow equations for the inverse
propagator \cite{CWRG,Tetradis,Aoki:2000wm,Pawlowski:2005xe} 
and therefore for $Z_\varphi,Z_\psi$. We define
\begin{equation}\label{20}
  \hat{\eta}_{\psi,\varphi}=\eta_{\psi,\varphi}/2k^2=-
  \partial\ln Z_{\psi,\varphi}/\partial k^2
\end{equation}
and obtain in our approximation
\begin{eqnarray}\label{ZRFormula}
  \hat{\eta}_\varphi
  &=& -\frac{ h_\varphi^2 Q_\psi}{2} 
  \frac{\partial}{\partial k^2}I^{(F)}_\varphi + \hat\eta_\varphi\hB.
\end{eqnarray}
Here the fermion loop integral
\begin{eqnarray}
  I^{(F)}_{\varphi} = \frac{1}{8\hat T^2} \int
  \frac{d^3 \hat q}{(2\pi)^3}\gamma \gamma_\varphi^{-3}[\tanh
  \gamma_\varphi -
  \gamma_\varphi \cosh^{-2}\gamma_\varphi]
\end{eqnarray}
involves
\begin{eqnarray}
  \gamma_\varphi &=& \frac{1}{2\hat T}\sqrt{(A_\psi 
    \hat q^2 +k^2-\sigmah)^2 + h_\varphi^2 \rho},\nonumber\\
  \gamma &=& \frac{1}{2\hat T}(A_\psi \hat q^2 +k^2-\sigmah).
\end{eqnarray}
In our truncation, the boson fluctuations contribute to
$\hat{\eta}_\varphi$ only in the superfluid phase
($\hat\eta_\varphi\hB(\rho=0) =0$) \cite{Diehl:2007th}.

Crucial quantities for the investigation of this problem are the
running of the Yukawa coupling $h_\varphi$,
\begin{equation}\label{21}
  \frac{\partial}{\partial k^2} h_\varphi^2  = (\hat\eta_\varphi 
  + 2\hat\eta_\psi) h_\varphi^2
  + 2h_\varphi^2\lambda_\psi Q_\psi J^{(F)}_{\varphi},
\end{equation}
with
\begin{eqnarray}
  J^{(F)}_\varphi &=& \frac{1}{16\hat T^2} \int
  \frac{d^3 \hat q}{(2\pi)^3}\gamma \gamma_\varphi^{-5}
  [(3\gamma^2 - \gamma_\varphi^2)\tanh\gamma_\varphi \nonumber\\
  &&\qquad+ \gamma_\varphi \cosh^{-2}\gamma_\varphi
  \big(-3\gamma^2 +\gamma_\varphi^2\nonumber\\
  &&\qquad\qquad +2\gamma_\varphi(\gamma_\varphi^2 - 
  \gamma^2)\tanh\gamma_\varphi\big)],
\end{eqnarray}
and the point-like four-fermion interaction
\begin{eqnarray}
  \frac{\partial}{\partial k^2} \lambda_\psi  &=& 2\hat
  \eta_\psi \lambda_\psi  +
  \lambda_\psi^2 Q_\psi(I^{(F)}_\varphi+I_{\lambda_\psi}),\\\nonumber
  I_{\lambda_\psi} &=&-\frac{1}{4\hat T^2} \int\frac{d^3 
    \hat q}{(2\pi)^3} \gamma \gamma_\varphi^{-1} \tanh 
  \gamma_\varphi   \cosh^{-2}\gamma_\varphi.
\end{eqnarray}
For both quantities, we neglect the contribution from the molecule
fluctuations. This is a valid approximation for the parameter ranges
for which we give quantitative results. In other ranges, the molecule
fluctuations give more dominant contributions, see \cite{Diehl:2007th}
and the discussion below.  For $\rho=0$, we note $J_\varphi\hF =
I_\varphi\hF$.

\section{Flow of the effective potential}
\label{flow}

Keeping now $\rho$ (and not $\hat\rho$) fixed, the flow of the
effective potential obeys,
\begin{equation}\label{23}
  \frac{\partial}{\partial k^2} u (\rho) = \zeta_F(\rho)
  + \zeta_B(\rho) +
  \hat\eta_\varphi \rho u'(\rho),
\end{equation}
with
\begin{eqnarray}\label{sigma}
  \zeta_F(\rho) &= &-Q_\psi\int\frac{d^3 \hat q}{(2\pi)^3}
  \Big(\frac{\gamma}{\gamma_\varphi} \tanh \gamma_\varphi - 1\Big),
  \nonumber\\
  \zeta_B(\rho) &=& Q_\varphi \int\frac{d^3 \hat q}{(2\pi)^3}
  \Big(\frac{\alpha + \kappa_2}{\alpha_\varphi}
  \coth \alpha_\varphi - 1\Big).
\end{eqnarray}
Here, we define
\begin{eqnarray}
  \alpha_\varphi&=&
  \frac{1}{2\hat{T}}\Big[(
  A_\varphi\hat{q}^2+2k^2+u')\nonumber\\
  &&(A_\varphi\hat{q}^2+2k^2+u'+2\rho u'')\Big]^{1/2},\nonumber\\
  \alpha &=& \frac{A_\varphi \hat{q}^2+2k^2  + u' }{2\hat T}, 
  \nonumber\\
  \kappa_2 &=& \frac{\rho u''}{2\hat T}~,~\kappa_3=
  \frac{\rho^2u'''}{2\hat{T}},
\end{eqnarray}
and the primes denote derivatives with respect to $\rho$. The
contribution of the boson fluctuations is computed in a basis where
the complex field $\varphi$ is written in terms of two real fields
with inverse bosonic propagator
\begin{eqnarray}
{\mathcal{P}}_\varphi  = \left(\!
\begin{array}{cc}
  {A_\varphi \hat q^2 +2k^2+ u' + 2\rho u''} & {-2\pi n\hat T} \\
  {2\pi n \hat T } & {A_\varphi \hat q^2 +2k^2+ u'}
\end{array}\!\right)\!.
\end{eqnarray}
Then, Eq.~(\ref{ExFlowEq}) evaluated for constant $\varphi$ becomes
\begin{eqnarray}
  \zeta_B(\rho) = \hat T Q_\varphi\mathrm{tr} \sum\limits_n \int
  \frac{d^3\hat q}{(2\pi)^3} 
  \left( \mathcal P_\varphi^{-1}
    -\textbf{1}\frac{1}{A_\varphi \hat{q}^2+2 k^2} \right)\!
  , 
\end{eqnarray}
in accordance with \cite{Gollisch02}. The subtraction of a
$\rho$-independent part renders the momentum integral ultraviolet
finite. For extensive numerical studies one would rather rely on
cutoff choices as in \cite{Diehl:2007th} which do not require any
ultraviolet subtraction, and minimise the numerical costs. However,
this is of no importance for the quantities computed in this paper --
the dependence of $A_\varphi$ on $\hat\sigma$ and $\hat T$ is
negligible, and simplicity of the approach is the more important
property.

In the symmetric phase (SYM), one has $\rho_0=0$, $u'(0) =
m_\varphi^2,u''(0)=\lambda_\varphi$, whereas in the presence of
spontaneous symmetry breaking (SSB) we use $\rho_0 >0, u'(\rho_0) =0,
u''(\rho_0) = \lambda_\varphi$.  Eq. (\ref{23}) yields in the
symmetric phase
\begin{equation}\label{mFlow}
  \frac{\partial}{\partial k^2}  m_\varphi^2 = h_\varphi^2 Q_\psi
  I^{(F)}_\varphi(\rho=0)
  -Q_\varphi I^{(B)}_\varphi (\rho=0)+  \hat\eta_\varphi m_\varphi^2.
\end{equation}
Here, we define the boson loop integral
\begin{eqnarray}
  I^{(B)}_\varphi&=&\frac{\lambda_\varphi}{2\hat{T}}
  \int\frac{d^3\hat{q}}{(2\pi)^3}
\label{A3mFlow}\\
&&\quad\times \left\{  \Big[(2\alpha+\kappa_2)u''+\alpha\rho
  u'''\Big] \frac{\alpha+\kappa_2}
  {\alpha_\varphi^2\sinh^2\alpha_\varphi}\right. \nonumber\\ 
&&\quad\quad\left.+\Big[\kappa_2(\kappa_2-\alpha)-\alpha\kappa_3\Big]u''
  \frac{\coth\alpha_\varphi}{\alpha_\varphi^3}\right\},\nonumber
\end{eqnarray}
such that
\begin{equation}\label{A3,1mFlow}
  I^{(B)}_\varphi (\rho=0)=  \frac{\lambda_\varphi}{\hat T}\int
  \frac{d^3 \hat q}{(2\pi)^3} \sinh^{-2} \alpha.
\end{equation}

The flow starts with a positive $m_\varphi^2$ for large $k$ and is
therefore in the symmetric regime. As $k$ decreases, $m_\varphi^2$
decreases according to Eq.~(\ref{mFlow}), with $\hat\eta_\varphi$
being positive. At high temperature, $m_\varphi^2$ stays positive for
all $k$ and the system is in the symmetric phase and ungapped for all
momenta. At temperatures below a pseudo-critical temperature
$T_{\text{p}}$, $m_\varphi^2$ hits zero at some critical
$k_{\text{c}}$. For $k<k_{\text{c}}$, the flow first continues in the
broken-symmetry regime with nonzero $\rho_0$. Whereas fermionic
fluctuations tend to increase $\rho_0$, molecule fluctuations cause
its depletion. For temperatures inbetween the critical temperature and
the pseudo-critical temperature, $T_{\text{c}}<T<T_{\text{p}}$,
molecule fluctuations eventually win out over the fermionic
fluctuations and $\rho_0$ vanishes again at smaller $k$; here, the
system is in the symmetric phase. Identifying the $k$ dependence of
the fermionic 2-point function with the momentum dependence, the
non-zero value of $\rho_0$ for finite $k$ can be associated with a
pseudo-gap \cite{Ranninger}. Below the critical temperature
$T_{\text{c}}$, $\rho_0$ stays finite for $k\to 0$, corresponding to
the superfluid phase with a truly gapped spectrum.

Molecule fluctuations are particularly important during those stages
of the flow where $\rho_0$ is non-zero, i.e., for $k<k_{\text{c}}$ and
below the pseudo-critical temperature $T_{\text{p}}$, see
\cite{Diehl:2007th}. In this work, we focus on the properties of the
flow in the symmetric regime, where $\rho_0=0$ for all values of $k$
under consideration. We still include molecule fluctuations for all
quantities that are associated with the effective potential $u(\rho)$,
in particular the molecule density and the molecule-molecule
scattering length, see Sects. \ref{density} and \ref{sec:scat}
below. But molecule fluctuations are neglected for the other running
couplings which are dominated by fermion fluctuations in the parameter
ranges considered here. 

\section{Solving the flow equation for zero temperature}
\label{solving}

Let us first concentrate on $T=0$ and $k^2-\sigmah >0$. We define
\begin{equation}\label{B5}
  K^2=k^2-\sigmah
\end{equation}
and employ $\partial/\partial k^2=\partial/\partial K^2$. Then the
flow equation for $u$ takes the explicit form
\begin{widetext}
\begin{eqnarray}
  \frac{\partial}{\partial K^2} u (\rho) =  \hat\eta_\varphi 
  \rho u' 
  &-&\frac{1}{ 2\pi^2}\int\limits_0^\infty dx x^2\Bigg\{Q_\psi
  \left(\frac{A_\psi x^2 +K^2}{\sqrt{(A_\psi x^2 +K^2)^2
        + h_\varphi^2\rho}} - 1\right) \label{T0flow}\\
  &&\qquad\qquad \qquad -Q_\varphi  \left(\frac{
      A_\varphi x^2+2k^2+u'+\rho u''}{\sqrt{(A_\varphi x^2+2k^2+u')
        (A_\varphi x^2+2k^2+u'+2\rho u'')}}
    - 1\right)\Bigg\}.\nonumber
\end{eqnarray}
\end{widetext}
For $\rho=0,T=0,\hat{\sigma}<0$, one also finds
\begin{equation}\label{A4a}
  I^{(F)}_\varphi=
  \frac{1}{16\pi}A^{-3/2}_\psi K^{-1}~,~\qquad I_{\lambda_\psi}=0,
\end{equation}
and, in our approximation, the contributions of the boson loops to
$\hat\eta_\varphi,
\partial h_\varphi^2/\partial k^2$ and $\partial \lambda_\psi
/\partial k^2$ vanish for $m_\varphi^2 >0$.

At $\hat T=0$ and $\hat\sigma<0$, we can use $Z_\psi=A_\psi=1$,
$Q_\psi=1$. This is an exact result \cite{Sachdev06} as long as all
propagators in the relevant diagrams have simple poles in the
imaginary $q_0$ plane. With
\begin{equation}\label{A4b}
\hat{\eta}_\varphi=\frac{h^2_\varphi}{64\pi}K^{-3},
\end{equation}
we find the coupled system of flow equations for the symmetric phase
\begin{eqnarray}\label{A4c}
\frac{\partial h^2_\varphi}{\partial K}&=&\frac{h^4_\varphi}{32\pi}K^{-2}+
\frac{h^2_\varphi\lambda_\psi}{4\pi},\nonumber\\
\frac{\partial\lambda_\psi}{\partial K}&=&\frac{\lambda^2_\psi}{8\pi}.
\end{eqnarray}
The solution of the last equation,
\begin{eqnarray}\label{A4d}
\lambda^{-1}_\psi&=&\lambda^{-1}_{\psi,\text{in}}+\frac{1}{8\pi}
\left(K_{\text{in}}-K\right)\nonumber\\
&=&\lambda^{-1}_{\psi,0}-\frac{K}{8\pi},
\end{eqnarray}
renders $\lambda_\psi$ almost independent of $K$ if $K\ll
K_{\text{in}}+8\pi/\lambda_{\psi, \text{in}}$.  Here, we assume
implicitly that $\lambda_{\psi,\text{in}}$ is not too much negative
such that $\lambda_\psi$ remains finite in the whole $k$ range of
interest. For positive $\lambda_\psi$, the self-consistency of the
flow requires an upper bound $\lambda_{\psi,0}/(8\pi) <
1/K_{\text{in}}$.

In contrast, for $\lambda_\psi=0$ the solution for $h^2_\varphi$,
\begin{equation}\label{A4e}
h_\varphi^{-2}=h^{-2}_{\varphi, \text{in}}+\frac{1}{32\pi}
\left(K^{-1}-K^{-1}_{\text{in}}\right),
\end{equation}
is dominated by small $K$.

For $\lambda_\psi\neq 0$, the flow of $h_\varphi$ is modified,
however, without changing the characteristic behaviour for $K
\rightarrow 0$. Indeed, in the limit $K\rightarrow 0$ the term
$\sim\lambda_\psi$ becomes subdominant for the evolution of
$h_\varphi^2$. The flow for the ratio $h^2_\varphi/K$ reaches a fixed
point $32\pi$. We can explicitely solve the system (\ref{A4c}) by
consecutive integrations. One obtains for $\hat{h}^2_\varphi=Z_\varphi
h^2_\varphi$
\begin{eqnarray}\label{A4f}
  \hat{h}^2_\varphi&=&\hat{h}^2_{\varphi,\text{in}}\exp
  \Bigg\{-\frac{1}{4\pi}\int\limits^{K_{
      \text{in}}}_{K}dx\lambda_\psi(x)\Bigg\}
  \nonumber\\
  &=&\hat{h}^2_{\varphi,\text{in}}\left(1-\hat c_0K_{
      \text{in}}\right)^2
  \left(1-\hat c_0K\right)^{-2}\nonumber\\
  &=& \hat h_{\varphi,0}^2\big(1 - \hat c_0K\big)^{-2}.
\end{eqnarray}
This yields for the wave function renormalisation
\begin{eqnarray}
  Z_\varphi &=& 1+\frac{1}{32\pi}\int\limits^{K^{-1}}_{
    K^{-1}_{\text{in}}}d(x^{-1})
  \hat{h}^2_\varphi(x^{-1})\nonumber\\
  &=&1 + \frac{\hat{h}^2_{\varphi,0}}{32\pi}\left[\frac{1}{K}
    \left(1 - \frac{\hat{c}_0}{1/K - \hat{c}_0}\right)
  \right.\label{45}\\
  &&\qquad\qquad \left.+ 2\hat{c}_0 \ln\left({\frac{1}{K} -
        \hat{c}_0}\right)
    - (K \to  K_{\text{in}})\right],
  \nonumber
\end{eqnarray}
with
\begin{eqnarray}
\hat{c}_0 = \frac{\lambda_{\psi,0}}{8\pi},\quad \frac{1}{K} >\hat{c}_0.
\end{eqnarray}
For $K \rightarrow 0$, the ratio
$\hat{h}^2_\varphi/\hat{h}^2_{\varphi,\text{in}}$ obviously reaches a
constant which depends on $\lambda_{\psi,\text{in}}, K_{\text{in}}$,
whereas $Z_\varphi$ diverges $\approx \hat{h}^2_{\varphi,0}/(32\pi
K)$, as is consistent with the fixed-point behaviour
$h^2_\varphi\approx 32\pi$.  Neglecting $\hat c_0$ (as always
appropriate for small enough $K$) and assuming a broad resonance $\hat
h^2_{\varphi,0}\gg 32\pi$, we arrive at the simple relation
\begin{equation}\label{b6a}
  Z_\varphi\approx\frac{\hat h^2_{\varphi,0}}{32\pi K}.
\end{equation}
In this limit, the factor $Q_\varphi$ appearing in the flow equations
is given by
\begin{equation}\label{b6b}
Q_\varphi=1-\frac{k^2}{2(k^2-\sigmah)}
\end{equation}
and approaches $1/2$ for $k^2\gg-\sigmah$. We note that in this range
the bosonic cutoff function $R^{(B)}_k=2Z_\varphi k^2$ is effectively
linear in $k$ and not quadratic.

We finally investigate the flow equation for $m_\varphi^2$ in the
symmetric phase. As we will see below, the boson loops (the last
contribution in Eq.~(\ref{T0flow})) vanish in our truncation. This
yields
\begin{equation}\label{mflowT0}
  \frac{\partial m^2_\varphi}{\partial K}=
  \frac{h^2_\varphi}{8\pi}+\frac{h^2_\varphi m^2_\varphi}{32\pi}K^{-2},
\end{equation}
or
\begin{equation}\label{50A}
  \frac{\partial \hat m^2_\varphi}{\partial K}
  =\frac{\partial}{\partial K}( Z_\varphi m^2_\varphi)
  =\frac{\hat h^2_\varphi}{8\pi}.
\end{equation}
The solution reads
\begin{eqnarray}\label{b6c}
  \hat m^2_\varphi&=&\frac{2M\bar{\mu}(B-B_0)}{\hat k^2}-2\sigmah 
  +\delta \hat\nu\\
  &&-\frac{\hat h^2_{\varphi,\text{in}}(1-
    \hat c_0 K_{\text{in}})^2}{8\pi\hat c_0}
  \left(\frac{1}{1-\hat c_0 K_{\text{in}}}-\frac{1}{1-
      \hat c_0 K}\right).\nonumber
\end{eqnarray}
The counterterm $\delta\hat\nu$ is determined from the condition $\hat
m^2_\varphi (B=B_0,\sigmah=0,k=0)=0$, such that (up to minor
corrections)
\begin{equation}\label{ZAB}
  \hat m^2_\varphi=\frac{2M\bar{\mu}(B-B_0)}{\hat k^2}
  +\frac{\hat h^2_{\varphi,0}}{8\pi}\frac{K}{1-\hat c_0K}-2\sigmah.
\end{equation}
In particular, for $k=0$ and $\sigmah<0$ one has $K=\sqrt{-\sigmah}$,
and Eq.~(\ref{ZAB}) yields the expression for $\hat m^2_\varphi$ in
the presence of all fluctuations
\begin{equation}\label{ZMM}
  \hat m^2_\varphi=\frac{2M\bar{\mu}(B-B_0)}{\hat k^2}+
  \frac{\hat h^2_{\varphi,0}\sqrt{-\sigmah}}{8\pi
    -\lambda_{\psi,0}\sqrt{-\sigmah}}-2\sigmah.
\end{equation}
At this point, we have the explicit solution of the flow equation for
$T=0$. In the present simple truncation, we expect that the
description of the system is qualitatively correct. Higher
quantitative precision will require a more extended truncation;
however, this is not the main emphasis of the present work which
rather concentrates on the structural properties related to the fixed
points.

\section{Flow equations and fixed points}
\label{flowandfixed}

\begin{figure}
\begin{minipage}{\linewidth}
\begin{center}
\setlength{\unitlength}{1mm}
\begin{picture}(85,55)
  \put (5,0){
    \makebox(80,50){
      \begin{picture}(80,50)
                \put(0,0){\epsfxsize80mm\epsffile{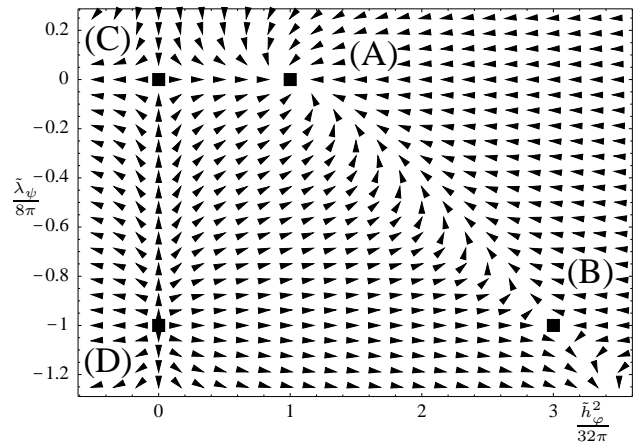}
        }
        \put(72,-2){$\frac{\tilde{h}_\varphi^2}{32\pi}$}
        \put(-3,28){$\frac{\tilde{\lambda}_{\psi}}{8\pi}$}
              \end{picture}
  }}
\end{picture}
\end{center}
\vspace*{-1.25ex} \caption{Location of fixed points (A)-(D) projected
  onto the plane which is spanned by the couplings
  $\tilde{\lambda}_{\psi}/(8\pi)$ and $\tilde{h}_\varphi^2/(32\pi)$.
  The arrows characterise the flow of the couplings towards the
  infrared. }
\label{fig:FPs}
\end{minipage}
\end{figure}

The explicit solution of the preceding section is indeed very useful
for verifying and explicitly demonstrating the general fixed-point
properties.  The following features of the flow equations hold in a
much wider context of different microscopic actions and different
cutoffs. The overall pattern of the flow is governed by the existence
of fixed points. Some of these fixed points may correspond to
particularly simple situations, being less relevant for the physics
under discussion. By contrast, the stability or instabilities of small
deviations from the various fixed points are much more important, as
they determine the topology of the flow in the space of coupling
constants, as demonstrated in Fig.~\ref{fig:FPs}. If the system is in
the vicinity of any of the different fixed points the number of
effective couplings needed for a description of the macrophysics
(beyond $T$) corresponds to the number of relevant directions.

A particularity of our system concerns a certain redundancy in the
description: a pointlike interaction can be described either by the
four-fermion coupling $\lambda_\psi$ or by a limiting behaviour of the
scalar exchange. Indeed, for $\bar h^2_{\varphi,\text{in}}\to\infty$
and fixed $\hat h^2_{\varphi, \text{in}}/\hat m^2_{\varphi,\text{in}}$
the molecule exchange interaction becomes effectively point-like.
Therefore, we may define an effective point-like coupling
\begin{equation}\label{b6d}
  \lambda_{\psi,\text{eff}}=\lambda_\psi-
  \frac{\hat h^2_\varphi}{\hat m^2_\varphi}=\lambda_\psi
  -\frac{h^2_\varphi}{m^2_\varphi}
\end{equation}
which describes the interaction in the zero-momentum limit. Combining
the flow equations (\ref{A4c}), (\ref{mflowT0}) for $\lambda_\psi,\hat
h^2_\varphi,\hat m^2_\varphi$, one obtains
\begin{equation}\label{b6e}
  \frac{\partial\lambda_{\psi,\text{eff}}}{\partial K}=
  \frac{\lambda^2_{\psi,\text{eff}}}{8\pi}.
\end{equation}
This is the same flow equation as for $\lambda_\psi$ (\ref{A4c}).

Next, we may consider the renormalised couplings in units of $k\hat k$
instead of $\hat k$. This will reveal the relevant fixed points for
the flow more clearly. We define, according to the scaling dimensions
of Eq.~(\ref{4}),
\begin{eqnarray}\label{b6f}
  \tilde{\lambda}_\psi&=&\lambda_\psi k~,\quad
  ~\tilde{\lambda}_{\psi,\text{eff}}=\lambda_{\psi,\text{eff}}k,\nonumber\\
  \tilde{h}_\varphi&=&\frac{h_\varphi}{\sqrt{k}}~,\quad
  ~\tilde{m}^2_\varphi=m^2_\varphi/k^2,\nonumber\\
  t&=&\ln k/k_{\text{in}},
\end{eqnarray}
and obtain the dimensionless flow equations
\begin{eqnarray}\label{YXX}
  \partial_t\tilde{\lambda}_\psi&=&\tilde{\lambda}_\psi
  +\frac{\tilde{\lambda}^2_\psi
    y}{8\pi}~,~\partial_t\tilde{\lambda}_{\psi,\text{eff}}
  =\tilde\lambda_{\psi,\text{eff}}
  +\frac{\tilde{\lambda}^2_{\psi,\text{eff}}y}{8\pi},\nonumber\\
  \partial_t\tilde{h}^2_\varphi&=&-\tilde{h}^2_\varphi+
  \frac{\tilde {h}_\varphi^4
    y^3}{32\pi}+\frac{\tilde{h}^2_\varphi
    \tilde{\lambda}_{\psi}y}{4\pi},\nonumber\\
  \partial_t\tilde m^2_\varphi&=&-2\tilde m^2_\varphi+\frac{\tilde
    h^2_\varphi y}{8\pi}+\frac{\tilde h^2_\varphi\tilde m^2_\varphi
    y^3}{32\pi},
\end{eqnarray}
with
\begin{equation}\label{YUV}
  y=\left(\frac{k^2}{k^2-\hat{\sigma}}\right)^{1/2}.
\end{equation}

Let us first consider $\hat{\sigma}=0$, or, more generally,
$k^2\gg-\hat{\sigma}$, such that $y=1$. In this case, we observe two
fixed points for $\tilde{\lambda}_\psi$, either
$\tilde{\lambda}_\psi=0$ or $\tilde{\lambda}_\psi=-8\pi$.  The
corresponding fixed points for $\tilde{h}^2_\varphi\neq 0$ are
\begin{eqnarray}\label{ZBC}
\begin{array}{lllll}
  \text{(A)}:\quad\tilde{\lambda}_\psi=0&,&\tilde{h}^2_\varphi
  =32\pi&,&\tilde{m}^2_\varphi=4\\
  \text{(B)}:\quad\tilde{\lambda}_\psi=-8\pi
  &,&\tilde{h}^2_\varphi=96\pi&,&\tilde{m}^2_\varphi=-12.\end{array}
\end{eqnarray}
(We will discuss later the fixed points with
$\tilde{h}_\varphi^2=0~,~\tilde{m}^2_\varphi=0$.) The fixed point (A)
is infrared stable for $\tilde{\lambda}_\psi$ and
$\tilde{h}^2_\varphi$ -- both couplings run towards their fixed-point
values as $k$ is lowered, as is visualised in Fig.~\ref{fig:FPs}. In
contrast, $\tilde{m}^2_\varphi$ is infrared unstable -- the detuning
$B-B_0$ corresponds to a relevant perturbation of the fixed point.
With $\tilde{h}^2_\varphi$ at the fixed-point value,
$\tilde{m}^2_\varphi$ deviates from its fixed point with an anomalous
dimension

\begin{equation}\label{b6g}
  \tilde{m}^2_\varphi=4+\delta\tilde{m}^2_{\text{in}}\frac{k_{\text{in}}}{k}.
\end{equation}
Comparing this with the explicit solution (\ref{ZAB}) for
$\lambda_\psi=0,\hat\sigma=0,\hat h_\varphi=\hat h_{\varphi,0}$,
namely
\begin{equation}\label{b6h}
\tilde{m}^2_\varphi=\frac{\tilde
  h^2_\varphi}{8\pi}+\frac{2M\bar\mu(B-B_0)}{Z_\varphi k^2\hat k^2}
\end{equation}
with Eq.~(\ref{45}),
\begin{equation}\label{b6i}
Z_\varphi\approx\frac{\hat h^2_{\varphi,0}}{32\pi k},
\end{equation}
we can identify
\begin{equation}\label{b6j}
  \delta m^2_\varphi=\frac{64\pi M\bar 
    \mu(B-B_0)}{\hat k^2\hat h^2_{\varphi,0}k_{\text{in}}}.
\end{equation}
We conclude that broad Feshbach resonances (large enough Yukawa
couplings) can be characterised by the fixed point (A), with $\tilde
h_\varphi$ and $\tilde\lambda_\psi$ being irrelevant and $\tilde
m^2_\varphi\sim B-B_0$ the relevant coupling.

For the second fixed point (B), $\tilde\lambda_\psi$ becomes a
relevant parameter. For $\tilde{h}^2_\varphi$, the fixed point
(\ref{ZBC}) with $\tilde h^2_\varphi=96\pi$ remains infrared
attractive, but the fixed point for $\tilde m^2_\varphi$ occurs for
negative $\tilde m^2_\varphi$ where our computation in the symmetric
phase is no longer valid.

Finally, we turn to the fixed points with $\tilde h^2_\varphi=0$. In
this case, $\tilde h^2_\varphi$ is always a relevant coupling and
increases as $k$ is lowered. The fixed point
\begin{equation}\label{b6k}
  \text{(C)}:\quad\tilde\lambda_\psi=0~,
  ~\tilde h^2_\varphi=0~,~\tilde m^2_\varphi=0
\end{equation}
is infrared attractive for $\tilde\lambda_\psi$. For this case of a
narrow Feshbach resonance, the crossover problem can be solved exactly
\cite{Diehl:2005an}, and perturbation theory around the exact solution
becomes valid for small $\tilde h^2_\varphi,\tilde\lambda_\psi$. The
coupling $\tilde m^2_\varphi$ is a second relevant parameter around
this fixed point. As one increases $\hat h^2_{\varphi,0}$, a crossover
to the fixed point (A) occurs \cite{Diehl:2005an}. The
``narrow-resonance fixed point'' (C) describes the limit of a combined
model with free fermions and free bosons, sharing the same chemical
potential. We emphasise that, for small $\tilde{h}_\varphi$, the
equivalent purely fermionic description typically has a nonlocal
interaction.

For the fourth fixed point of our system,
\begin{equation}\label{b6l}
  \text{(D)}:\quad\tilde\lambda_\psi=-8\pi~,~\tilde h^2_\varphi=0~,
  ~\tilde m^2_\varphi=0,
\end{equation}
all three parameters are relevant. This point corresponds again to
strong attractive interactions between the fermionic atoms. The flow
away from this fixed point for non-vanishing $\tilde h^2_\varphi/\tilde
m^2_\varphi$ can be characterised by the flow of the ratio $(y=1)$
\begin{equation}\label{b6m}
\partial_t\left(\frac{\tilde h^2_\varphi}{\tilde m^2_\varphi}\right)=
\left(1+\frac{\lambda_\psi}{4\pi}\right)
\frac{\tilde h^2_\varphi}{\tilde m^2_\varphi}-\frac{1}{8\pi}
\left(\frac{\tilde h^2_\varphi}{\tilde m^2_\varphi}\right)^2
\end{equation}
which does not have a fixed point for positive $\hat h^2_\varphi,\hat
m^2_\varphi$. Formally, the flow runs for negative $\hat m^2_\varphi$
towards the fixed point (B) with $\tilde h^2_\varphi/\tilde
m^2_\varphi=-8\pi$.  Actually, it may be possible to consider
$\lambda_\psi$ as a redundant parameter using partial bosonisation and
rebosonisation during the flow \cite{Gies01}, see also
\cite{Pawlowski:2005xe}. Then, the fixed points with
$\tilde{\lambda}_{\psi}=-8\pi$ may again be associated with broad
Feshbach resonances. We show the four fixed points A, B, C, D and the
infrared flow of the couplings $\tilde{h}_\varphi^2$ and
$\tilde{\lambda}_\varphi$ in Fig.~\ref{fig:FPs}

We should emphasise that the inclusion of the omitted contributions
from boson fluctuations to the flow of $\tilde\lambda_\psi$ could
result in corrections $(y=1)$,
\begin{equation}\label{b6n}
  \partial_t\tilde\lambda_\psi=\tilde\lambda_\psi+\frac{
    \tilde\lambda^2_\psi}{8\pi}+\frac{c_1}{32\pi}\tilde h^4_\varphi+
  \frac{c_2}{32\pi}\tilde\lambda_\psi\tilde h^2_\varphi,
\end{equation}
such that the flow of $\tilde\lambda_\psi$ remains no longer
independent of $\tilde h^2_\varphi$. This will change the precise
location of fixed points (A) and (B). We expect that the qualitative
characteristics of the fixed point (A) remain unchanged, whereas the
fate of the fixed point (B) is less clear. We may consider the flow of
the ratio $\tilde\lambda_\psi/\hat h^2_\varphi$,
\begin{eqnarray}
\partial_t\left(\frac{\tilde\lambda_\psi}{\tilde h^2_\varphi}\right)
&=&
2\left(\frac{\tilde\lambda_\psi}{\tilde h^2_\varphi}\right)
\label{b6o}\\
&&+
\frac{\tih^2_\varphi}{32\pi}
\left[c_1+(c_2-1)\frac{\tila_\psi}{\tih^2_\varphi}
-4\left(\frac{\tilde\lambda_\psi}{\hpn^2}\right)^2\right]. \nonumber
\end{eqnarray}
For any possible fixed point, one has
\begin{eqnarray}\label{b6p}
\hpn^2&=&32\pi\left(1-\frac{\tilde\lambda_\psi}{4\pi}\right) \quad
\Rightarrow\quad 
\frac{\hpn^2}{32\pi}=\left(1+8\frac{\tilde\lambda_\psi}{\hpn^2}\right)^{-1}
\!\!\!\!\!\!, \qquad
\end{eqnarray}
and therefore obtains the fixed point condition for
$Q=\tilde\lambda_\psi/\hpn^2$ as
\begin{equation}\label{b6q}
c_1+(c_2+1)Q+12Q^2=0.
\end{equation}
In general, this quadratic equation has two distinct solutions, where
the larger value of $Q$ is infrared stable and corresponds to (A),
while the smaller value of $Q$ is unstable and corresponds to (B).

We finally include the effect of a nonzero negative chemical potential
$\sigmah$. As soon as $k^2$ becomes smaller than $\-\sigmah$, the
parameter $y$ (\ref{YUV}) rapidly goes to zero. Then, only the first
term in the flow equations (\ref{YXX}) matters. In this range of $k$,
the couplings $\lan,\hpn^2,\tilde m^2_\varphi$ remain constant. A
negative $\sigmah$ acts as an additional infrared cutoff, such that the
flow is effectively stopped for $k^2<-\sigmah$. This yields a simple
rough picture: the couplings $\hat\lambda_\psi,\hat h^2_\varphi,\hat
m^2_\varphi$ flow according to Eq.~(\ref{YXX}) for $k^2>-\sigmah$
until the flow stops when $k^2<-\sigmah$.

\begin{figure}
\begin{minipage}{\linewidth}
\begin{center}
\setlength{\unitlength}{1mm}
\begin{picture}(85,108)
      \put (0,0){
     \makebox(80,49){
     \begin{picture}(80,49)
            \put(0,0){\epsfxsize80mm\epsffile{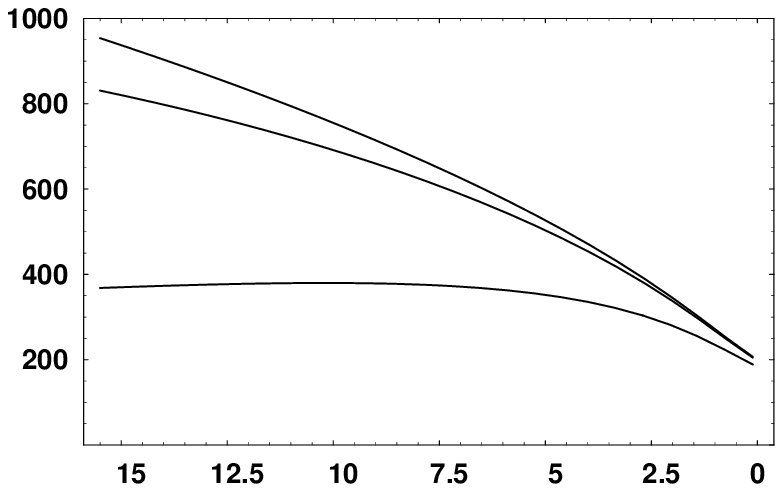}
}
      \put(67,-2){$k^2$}
      \put(68,40){$h_\varphi^2$ }
          \put(-3,1){(b)}
      \end{picture}
      }}
      \put (0,54){
    \makebox(80,49){
    \begin{picture}(80,49)
          \put(0,0){\epsfxsize80mm\epsffile{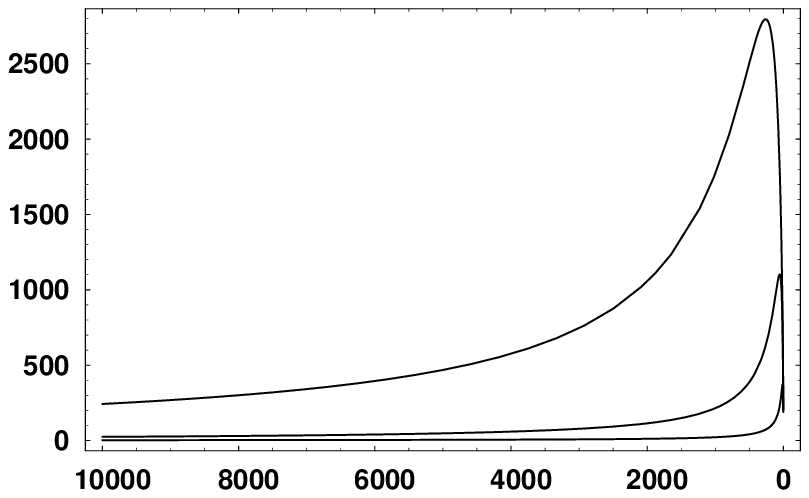}
}
      \put(67,-3){$k^2$}
       \put(12,40){$h_\varphi^2$ }
      \put(-3,1){(a)}
      \end{picture}
      }}
\end{picture}
\end{center}
\vspace*{-1.25ex} \caption{Sensitivity of the renormalised Yukawa
  coupling $h_\varphi$ to the two-body Yukawa coupling
  $\hat{h}_{\varphi,0}$ at fixed $\hat\lambda_{\psi,0}$: (a) UV and
  (b) IR flow of the Yukawa coupling. We use
  $\hat{k}=k_{\text{F}}=1\text{eV}$, $a_{\text{bg}}\hat{k}= \hat
  \lambda_{\psi,0}/(8\pi)= 0.38$ as appropriate for $\lit$ and $\hat
  T=0.5$, $c^{-1}=1$.  The different curves correspond to the two-body
  value of the Yukawa coupling $\hat{h}_{\varphi,0}^2 =3.72 \cdot
  (10^3,10^4,10^5)$ (from top to bottom), where the last value
  corresponds to $\lit$ while the first one is comparable to that of
  $\kal$. Universality with respect to the value of the two-body
  Yukawa coupling is very strong even for a comparatively large
  (fixed) $\hat\lambda_\psi$.} \label{hpUniversality}
\end{minipage}
\end{figure}

\begin{figure}
\begin{minipage}{\linewidth}
\begin{center}
\setlength{\unitlength}{1mm}
\begin{picture}(85,108)
      \put (0,0){
     \makebox(80,49){
     \begin{picture}(80,49)
            \put(0,0){\epsfxsize80mm\epsffile{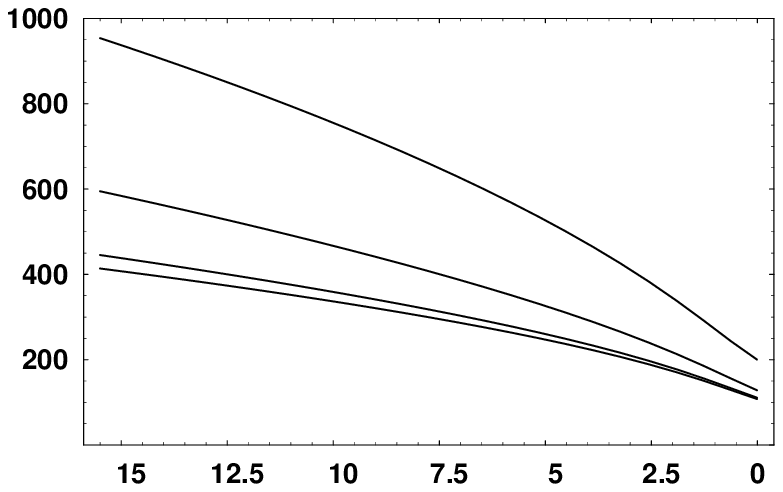}
}
      \put(67,-2){$k^2$}
      \put(68,40){$h_\varphi^2$ }
          \put(-3,1){(b)}
      \end{picture}
      }}
      \put (0,54){
    \makebox(80,49){
    \begin{picture}(80,49)
          \put(0,0){\epsfxsize80mm\epsffile{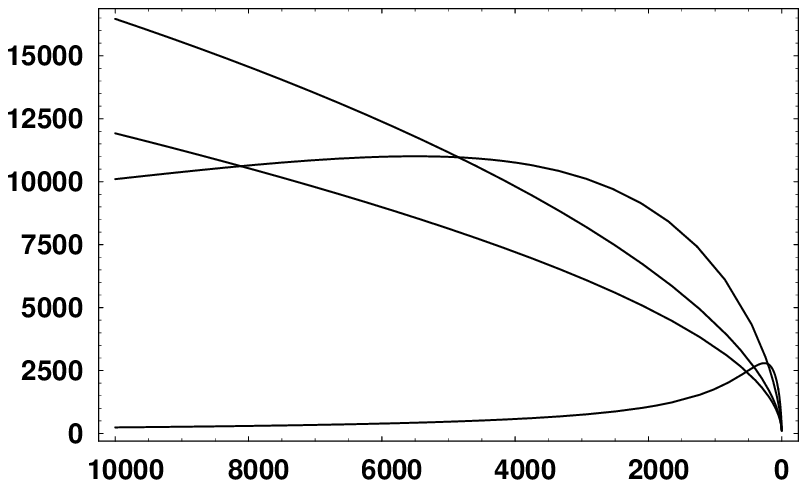}
}
      \put(67,-3){$k^2$}
       \put(12,40){$h_\varphi^2$ }
      \put(-3,1){(a)}
      \end{picture}
      }}
\end{picture}
\end{center}
\vspace*{-1.25ex} \caption{Sensitivity of the renormalised Yukawa
  coupling $h_\varphi$ to the two-body background scattering length
  $\hat\lambda_\psi$ for fixed two-body Yukawa coupling
  $\hat{h}_{\varphi,0}=3.72 \cdot 10^5 $, $\hat{k} =k_{\text{F}}=1
  \mathrm{eV}$, as appropriate for $\lit$: (a) UV and (b) IR flow of
  the Yukawa coupling. We plot curves for $a_{\text{bg}}\hat{k}= \hat
  \lambda_\psi/(8\pi)= -0.38 \cdot (10^{-3},10^{-2},10^{-1},1)$ (the
  last value corresponds to $\lit$) and $\Tn =0.5$, $c^{-1}= 1$. The
  uppermost line in (b) has the largest $a_{\text{bg}}$ and starts in
  (a) as the lowest line. }
\label{hpUniversality2}
\end{minipage}
\end{figure}

The fixed points are relevant not only for $T=0$. We demonstrate the
influence of the fixed point (A) on the flow of the Yukawa coupling
for $\hat T=0.5$ and $c=1$ in Figs. \ref{hpUniversality},
\ref{hpUniversality2}. In Fig.~\ref{hpUniversality}, we observe that
the renormalised Yukawa coupling $h_\varphi$ at small $k$ becomes
almost independent of its initial value $\hat{h}_{\varphi,0}^2$ if the
latter is large. Figure \ref{hpUniversality2} demonstrates the
influence of the background scattering length $a_{\text{bg}}$ for a
fixed value of $a$, again for $\hat T=0.5$ and $c=1$.

\section{Density and condensate fraction}
\label{density}

Let us next compute the density
\begin{equation}\label{DD1}
\hat n=\frac{n}{\hat k^3}=-\frac{\partial u_0}{\partial\sigmah},
\end{equation}
where the $\sigmah$ derivative will be taken at fixed $\rho$ and
$k$. This is possible, since for the minimum $du_0/d\sigma=\partial
u/\partial\sigma_{|\rho_0}+(\partial
u/\partial\varphi)(\partial\varphi_0/\partial\sigma) =\partial
u/\partial\sigma_{|\rho_0}$. Before doing a more detailed calculation,
we approximate the potential by
\begin{equation}\label{b6r}
u(\hat \rho,\sigmah)=\bar u(\sigmah)+\hat m^2_\varphi(\sigmah)\hat
\rho+\frac12\hat\lambda_\varphi\hat\rho^2 
\end{equation}
with $\lambda_\varphi$ independent of $\sigmah$. The density then receives
contributions from unbound atoms and molecules $(\hat
n_{\text{F}}+2\hat n_{\text{M}})$ and atoms in the condensate $(\hat
n_{\text{C}})$ for $\rho_0\neq0$,
\begin{equation}\label{b6s}
  \hat n_{\text{F}}+2\hat n_{\text{M}}=-\frac{\partial\bar u}{
  \partial\sigmah}~,~
\qquad 
n_{\text{C}}=-\frac{\partial\hat m^2_\varphi}{\partial
  \sigmah}\hat\rho_0.
\end{equation}
For an estimate of the condensate fraction $\Omega_{\text{C}}=\hat
n_{\text{C}}/\hat n$, we use (up to small corrections involving
$K_{\text{in}}$)
\begin{equation}
  \frac{\partial\hat m^2_\varphi}{\partial\sigmah}
  = -2 Z_\varphi \label{MMQ} 
  \left[1+ \mathcal{O}\left( \hat{c}_0 K\, \ln \frac{1}{K} \right) \right],
\end{equation}
where we have used \eq{b6a} and \eq{ZMM}. For small $K$, the term
$-2Z_\varphi$ dominates and we conclude that the condensate fraction
is given by the renormalised order parameter $\rho_0$,
\begin{equation}\label{b6t}
  \Omega_{\text{C}}=\frac{2\rho_0}{\hat n}.
\end{equation}
The above formula is one viable definition of the condensate
fraction. In general, it still needs to be related to the definition
of a condensate fraction as measured in a particular experiment.

The density receives contributions from atoms and molecules with
different momenta. In order to describe each momentum mode accurately,
we first compute the change of the density with $k$,
\begin{equation}\label{DDF}
  \frac{d}{dk^2}\hat n=-\frac{d}{dk^2}
  \frac{\partial u_0}{\partial\hat\sigma}=
  -\frac{\partial}{\partial k^2}
  \left.\frac{\partial u_0}{\partial\hat\sigma}\right|_{\rho_0}-
  \left.\frac{\partial^2 u}{\partial\rho\partial\hat\sigma}
  \right|_{\rho_0}
  \frac{d\rho_0}{dk^2},
\end{equation}
and then integrate from $k_{\text{in}}$ to $k=0$. The r.h.s. will be
dominated by momenta $q^2\approx k^2$. The first term in
Eq.~(\ref{DDF}) is given by the $\hat\sigma$ derivative of
Eq.~(\ref{sigma}). The different contributions have an approximate
interpretation in terms of unbound atoms, molecules and the condensate
density 
\ba
\label{NNY} \frac{d\hat
  n_{\text{F}}}{dk^2}&=&-\frac{\partial}{\partial\hat\sigma}
\zeta_F(\rho_0)~,~\qquad 2\frac{d\hat n_{\text{M}}}{dk^2}=-
\frac{\partial}{\partial\hat\sigma}\zeta_B(\rho_0),\nonumber\\
\frac{d\hat
  n_{\text{C}}}{dk^2}&=&-\left.\frac{\partial^2u}{\partial\rho
  \partial\hat\sigma}\right|_{\rho_0}
\left(\frac{d\rho_0}{dk^2}+\hat\eta_\varphi\rho_0\right).  \ea
For example, for $\rho_0=0$ one rediscovers the integral over the
usual Fermi distribution ($Q_\psi=1)$
\ba\label{NNZ} \frac{d\hat
  n_{\text{F}}}{dk^2}&=&\frac{\partial}{\partial k^2}\zeta_F(0)\\
&=&\frac{\partial}{\partial k^2}\int\frac{d^3\hat
  q}{(2\pi)^2}\frac{2}{\exp\{(A_\psi\hat q^2+k^2-\hat\sigma)/\hat
  T\}+1}.\nonumber \ea
In \eq{NNZ} we have traded the $\hat\sigma$ derivative for a $k^2$
derivative, since $\gamma$ depends only on the combination
$k^2-\hat\sigma$. As long as $A_\psi$ depends only on
$K^2=k^2-\hat\sigma$, the $k^2$ derivative also acts on $A_\psi$, and
Eq.~(\ref{NNZ}) can be integrated,
\ba\label{NCC} \hat n_{\text{F}}&=&\int\frac{d^3\hat
  q}{(2\pi)^3}\frac{2}{\exp\{(A_\psi\hat q^2-\hat\sigma)/\hat
  T\}+1}\nonumber\\ &&+\hat
n_{\text{F}}(k_{\text{in}})-\zeta_F(\rho=0,k_{\text{in}}).  \ea
We may evaluate the initial value $\hat n_{\text{F}}(k_{\text{in}})$
in perturbation theory, finding both quantities being exponentially
suppressed, $\hat
n_{\text{F}}(k_{\text{in}})=\zeta_F(\rho=0,k_{\text{in}})\approx 0$.
In our approximation, $(A_\psi=1)~\hat n_{\text{F}}$ is simply the
density of a free gas of fermionic atoms as long as $\rho_0=0$. In the
presence of a condensate $(\rho_0\neq 0)$, we typically find a flow
where $\rho_0(k)=0$ for $k>k_c$, such that the flow of $\hat
n_{\text{F}}(k)$ remains unchanged in this range.  On the other hand,
the contribution from modes with $q^2<k^2_c$ will be suppressed by the
presence of a gap $\Delta=h_\varphi\sqrt{\rho_0}$ in the propagator.
We emphasise that the derivative $\partial/\partial k^2$ in
Eq.~(\ref{NNZ}) does not act on $\rho_0$, since the $\hat\sigma$
derivative in Eq.~(\ref{NNY}) is taken at fixed $\rho$. Therefore the
flow for $\hat n_{\text{F}}$ has to be integrated numerically if
$\rho_0\neq 0$.

For the computation of the molecule density $\hat n_{\text{M}}$, we
need $\partial\zeta_B/\partial\hat\sigma$. Let us concentrate on
$\rho_0=0$ and neglect the $\hat\sigma$ dependence of $Q_\varphi$ and
$A_\varphi$, such that
\begin{eqnarray}
  \frac{\partial\zeta_B}{\partial\hat\sigma}&=&2Q_\varphi
  \int\frac{d^3\hat
    q}{(2\pi)^3}\frac{\partial}{\partial\hat\sigma}\nonumber\\
  &&\Big(\exp\big\{(A_\varphi\hat q^2+2k^2+m^2_\varphi)/\hat
  T\big\}-1\Big)^{-1}\nonumber\\ &=&Q_\varphi\frac{\partial
    m^2_\varphi}{\partial\hat\sigma}\frac{\partial}{\partial k^2}
  \int\frac{d^3\hat q}{(2\pi)^3}\label{b6u}\\
  &&\times\left(\exp\{(A_\varphi\hat q^2+2k^2+m^2_\varphi)/\hat
    T\}-1\right)^{-1}.  \nonumber
\end{eqnarray}
We next use Eq.~(\ref{MMQ}) 
\begin{equation}\label{NDA}
  \frac{dm^2_\varphi}{d\hat\sigma}\approx-2-
  m^2_\varphi\frac{\partial}{\partial\sigmah}\ln Z_\varphi.  
\end{equation}
The second term can be approximated for small $K$ by
\be
\label{b6v}
m^2_\varphi\frac{\partial}{\partial\sigmah}\ln
Z_\varphi\approx\frac{m^2_\varphi}{2(k^2-\sigmah)}.
\ee
 If we neglect this term and set $Q_\varphi=1$ we find the
intuitive formula
\be
\label{NAA} \hat
n_{\text{M}}=\int\frac{d^3q}{(2\pi)^3}\frac{1}{\exp\big\{(A_\varphi\hat
q^2+m^2_\varphi)/\hat T\big\}-1}.
\ee
This is the expression for a free boson gas. What is crucial,
however, is the appearance of the inverse propagator for the {\em
renormalised} bosonic field $\varphi$ instead of the microscopic or
``bare'' molecule field $\hat\varphi$. The propagators for the
renormalised and bare fields are related by a relative factor
$Z_\varphi$, such that we find for the density of bare molecules
\be\label{b6w} \hat
n_{\text{M},\text{bare}}=Z^{-1}_\varphi\hat n_{\text{M}}.
\ee
As a consequence, one may have a substantial molecule fraction $2\hat
n_{\text{M}}/\hat n$ even if the density of bare molecules is tiny.
These features reproduce the results from a solution of the
Schwinger-Dyson equations \cite{Diehl:2005an} where a bare molecule
density in accordance with the experimental result of Partridge \emph{et al.}
\cite{Partridge05} has been found. It is one of the important
advantages of our functional renormalisation group approach that it
accounts for the distinction between renormalised and bare molecules
in a very direct and straightforward manner \cite{Diehl:2007th}.
Within a Hamiltonian formulation, this issue is related to the mixing
between open-channel and closed-channel atoms, and its correct
treatment is crucial for a quantitatively reliable description of the
crossover. We emphasise that the renormalised molecule field plays an
important role even on the BCS side of the crossover. The composite
bosons are crucial effective degrees freedom, even though the
microscopic theory can be very well approximated by a point-like
interaction between fermionic atoms without any reference to
molecules. On the BCS side, the renormalised field $\varphi$ describes
Cooper pairs. Nevertheless, one never needs these physical
interpretations explicitly, since the density only involves the
$\sigmah$ dependence of the potential at its minimum, which is a
well-defined quantity.

We finally turn to the condensate density. If we can approximate the
$\sigmah$ dependence of $(u-\bar{u})$ by a term $-2\sigmah\rho$, as
suggested by Eq.~(\ref{NDA}), we infer 
\be\label{b6x}
 \frac{d\hat
n_{\text{C}}}{dk^2}=2\frac{d\rho_0}{dk^2}
+2\hat\eta_\varphi\rho_0=2Z_\varphi\frac{d}{dk^2}\frac{\rho_0}{Z_\varphi}. 
\ee 
(This approximation needs to hold only in the range of $k$ where
$\rho_0$ differs from zero.) The dominant flow of $\hat n_{\text{C}}$
typically arises from a region where the $k$ dependence of $Z_\varphi$
is already sub-leading, resulting in $\hat n_{\text{C}}\approx
2\rho_0$, as found above. In practice, all these various
approximations of our analytical discussion need not to be made, since
it is sufficient to follow the flow of $\hat n(k)$ numerically,
starting from an initial value $\hat n(k_{\text{in}})\approx 0$ and
extracting the physical density for $k=0$.

\section{Vacuum and two-atom scattering}
\label{vacuum}

It is one of the advantages of our method that it can access
simultaneously the many-body physics of a gas in thermal equilibrium
and the two-body physics of atom scattering and molecular
binding. Indeed, the two-body physics describes excitations above the
vacuum. In turn, the vacuum and the properties of its excitations
obtain in our formalism simply by taking the limit of vanishing
density and temperature. (More precisely, the limit should be taken
such that the ratio $\hat T/\hat n^{2/3}$ is large enough that no
condensate occurs.) The scattering cross section between two atoms can
then be directly inferred from the Yukawa coupling and the propagator
of the molecule field in the vacuum.

For $T=0$, the condition of zero density requires $\sigmah\leq 0$ in
order to ensure $\hat n_{\text{F}}=0$, cf. Eq.~(\ref{NCC}). On the
other hand, we infer from Eq.~(\ref{NAA}) that $m^2_\varphi\geq 0$ is
needed for $\hat{n}_{\text{M}}=0$, whereas a vanishing condensate
density requires $\rho_0=0$. The vacuum is the state that is reached
as density and temperature approach zero from above. It therefore
corresponds to the boundary of the region where $\sigmah\leq 0,
m^2_\varphi\geq 0$. There are two branches of vacuum states (for a
more detailed discussion, see \cite{Diehl:2007tg}). The first has a
negative $\sigmah =\sigmah_{\text{A}}<0$ and $m^2_\varphi=0$. In this
case, the single-atom excitations have a gap $-\sigmah_{\text{A}}>0$
which corresponds to half the binding energy of the stable molecules,
\be\label{b6y} \epsilon_{\text{M}}=\hat k^2\sigmah_{\text{A}}/M~,~\hat
\epsilon_{\text{M}}=2\sigmah_{\text{A}}.  \ee
We identify this state with the ``molecule phase'' where stable
molecules exist in the vacuum. This phase is realized for $B<B_0$,
with $\epsilon_{\text{M}}$ or $\sigmah_{\text{A}}$ being a function of
$B$ that vanishes for $B\to B_0$. The other branch corresponds to
$\sigmah=0,m^2_\varphi>0$. This ``atom phase'' of the vacuum is
realized for $B>B_0$, with $m^2_\varphi$ a function of $B$ vanishing
for $B\to B_0$. In the atom phase the ``binding energy''
$\epsilon_{\text{M}}=\hat k^2m^2_\varphi/(2M)$ is positive and the
``molecules'' correspond to unstable resonances. We observe a
continuous phase transition between the molecule and atom phase
\cite{Diehl:2005an,Diehl:2007th} for $m^2_\varphi=0,\sigmah=0$,
corresponding to the location of the Feshbach resonance at $B=B_0$.
This fixes $\delta\nu$ in the initial value of $m^2_\varphi$
(\ref{14}) by the requirement that for $\nu=0$ the mass term
$m^2_\varphi(\sigmah)$ vanishes precisely for $\sigmah=0$ if $T=0$.

The vacuum state can be used to fix the parameters of our model by
direct comparison to experimentally measured binding energies and
cross sections. Let us first consider the molecular binding energy
$\epsilon_{\text{M}}(B)$ that can be computed from Eq.~(\ref{ZMM}) by
requiring $\hat m^2_\varphi=0$, i.e.,
\begin{eqnarray}\label{EE}
  \epsilon_{\text{M}}(B)&=&\frac{\hat\sigma_{\text{A}}(B)\hat{k}^2}{M}\\
  &=&\bar{\mu}(B-B_0)\nonumber\\
  &&+\frac{\hat{k}\hat{h}^2_{\varphi,0}}{16\pi\sqrt{M}}
  \sqrt{|\epsilon_{\text{M}}|}
  \left(1-\frac{\lambda_{\psi,0}\sqrt{M|
        \epsilon_{\text{M}}|}}{8\pi\hat{k}}\right)^{-1}.\nonumber
\end{eqnarray}
In the vicinity of the Feshbach resonance (small
$|\epsilon_{\text{M}}|)$ the binding energy depends quadratically on
$B$
\begin{equation}\label{FF}
  \epsilon_{\text{M}}=-\frac{(16\pi)^2M\bar{\mu}^2(B_0-B)^2}{
    \hat{k}^2\hat{h}^4_{\varphi,0}}.
\end{equation}
Using
\begin{eqnarray}\label{GG}
  a_0&=&\frac{\lambda_{\psi,0}}{8\pi \hat{k}}~,~\bar{h}^2_{\varphi,0}=
  \frac{\hat{k}\hat{h}^2_{\varphi,0}}{4M^2}~,~\nonumber\\
  \tilde a&=&-\frac{\bar{h}^2_{\varphi,0}M}{4\pi\Big[\bar{\mu}(B-B_0)
    -\epsilon_{\text{M}}\Big]}+a_0
\end{eqnarray}
Eq.~(\ref{EE}) reads
\begin{equation}\label{HH}
\sqrt{M|\epsilon_{\text{M}}|}=\frac{1}{\tilde a}.
\end{equation}

The scattering length for the fermionic atoms is determined by the
total cross section in the zero-momentum limit, $\sigma = 4\pi a^2$,
such that
\begin{eqnarray}\label{FermScatt}
  a = \frac{M}{4\pi} {\bar{\lambda}_{\psi,\text{eff}}} = 
  \frac{\lambda_{\psi,\text{eff}}}{8\pi \hat k}.
\end{eqnarray}
More details are described in the appendix. For the atom phase
($\epsilon_{\text{M}} >0$), one has \cite{Diehl:2005an}
\begin{eqnarray}
  \lambda_{\psi,\text{eff}} = \lambda_{\psi,0} - \frac{\hat
    h_{\varphi,0}^2}{\hat m_\varphi^2(\sigmah = 0)} 
  = \lambda_{\psi,0}+\lambda_{\psi,\text{R}}.
\end{eqnarray}
For the molecule phase, the effective interaction depends on the
energy $\omega$ of the exchanged molecule
\begin{eqnarray}\label{60}
  \lambda_{\psi,\text{R}}(\omega=0,\vec{p}=0)  &=&
  - \frac{\hat h_{\varphi}^2(\sigmah_{\text{A}})}{2
    Z_{\varphi}(\sigmah_{\text{A}})\sigmah_{\text{A}}}, \nonumber\\
  \lambda_{\psi,\text{R}}(\omega=-2 \sigmah_{\text{A}},\vec{p}=0)  &=&
  - \frac{\hat h_{\varphi}^2(\sigmah_{\text{A}})}{4
    Z_{\varphi}(\sigmah_{\text{A}})\sigmah_{\text{A}}}.
\end{eqnarray}
We find for the atom phase
\begin{eqnarray}
  a = a_0 - \frac{\bar{h}_{\varphi,0}^2 M }{4\pi\bar\mu (B-B_0)}  = a_0
  + a_{\text{res}}, 
\end{eqnarray}
which agrees with $\tilde{a}$ (\ref{GG}) for $\epm =0$. Eq. 
(\ref{HH}) therefore relates the binding energy in the molecule phase
to the scattering length in the atom phase.

The value of $\hat h_{\varphi,0}$ can now be extracted from the
resonant part, $a_{\text{res}}$, whereas $\lambda_{\psi,0}$ follows
from the $B$-independent ``background scattering length'' $a_0
=a_{\text{bg}}$ which has been measured as
$a^{\text{(Li)}}_{\text{bg}}= -1420 a_{\text{B}}
~,~a^{\text{(K)}}_{\text{bg}}= 174 a_{\text{B}}$ ($a_{\text{B}}$ the
Bohr radius), or, expressed in units $\hbar = c = k_{\text{B}} =1$
($a_{\text{B}} = 2.6817 \cdot 10^{-4} \mathrm{eV}^{-1}$),
$a^{\text{(Li)}}_{\text{bg}}= -0.38
\mathrm{eV}^{-1}~,~a^{\text{(K)}}_{\text{bg}}= 4.67\cdot 10^{-2}
\mathrm{eV}^{-1}$.  For $\hat{k}=1\mathrm{eV}$ corresponding to a
density $n = 4.4 \cdot 10^{12} \mathrm{cm}^{-3}$, we find the
dimensionless expressions for $\lit$ and $\kal$:
\begin{eqnarray}\label{A4g}
  \hat{h}^{2\text{(Li)}}_{\varphi,0}&=&3.72\cdot 10^5~,\quad ~
  \hat{h}^{2\text{(K)}}_{\varphi,0}=6.1\cdot10^3,\\\nonumber
  \lambda_{\psi,0}^{\text{(Li)}}&=& - 9.55  ~,\quad~
  \lambda_{\psi,0}^{\text{(K)}}= 1.17.
\end{eqnarray}
From these values, one can compute the initial values
$\lambda_{\psi,\text{in}},h^2_{\varphi,\text{in}}$. For $\lit$, we use
the values
\begin{equation}
  k_{\text{in}} = 10^{3}, 
  \quad \hat{h}^{2\text{(Li)}}_{\varphi,\text{in}} = 2.56 ,\quad 
  \hat{\lambda}^{\text{(Li)}}_{\psi,\text{in}}
  =-2.51\cdot 10^{-2},
\end{equation}
whereas for $\kal$ we take
\begin{equation}
  k_{\text{in}} = \sqrt{300}, 
  \quad\hat{h}^{2\text{(K)}}_{\varphi,\text{in}} = 1.67\cdot 10^{5},\quad 
  \hat{\lambda}^{\text{(K)}}_{\psi,\text{in}} = 6.14.
\end{equation}
For $\kal$, we observe large values of $\hat{h}_{\varphi,\text{in}}^2$
and $\hat\lambda_{\psi,\text{in}}$. Similar values for $\lit$ are
observed in the corresponding $K$ range, e.g.
$\hat{h}_{\varphi,\text{in}}^2(K^2 = 300) = 6490,
\lambda_\psi^{\text{(Li)}} (K^2 = 300) = 2024$. The broad Feshbach
resonances indeed describe strongly coupled systems.  In our
approximation, the solutions of the flow equations for
$h^2_\varphi,\lambda_\psi$ and $Z_\varphi$ do not depend on
$m^2_\varphi$ provided $m^2_\varphi$ remains positive during the flow.

\section{Scattering length for molecules}
\label{sec:scat}

So far, we have only discussed the contributions of the molecule
fluctuations to $\partial u/\partial\sigmah$ in order to extract the
molecule density $\hat n_{\text{M}}$. In this section, we discuss
their influence on the effective potential for $T=0$ more
systematically. In particular, we will extract the scattering length
for molecule-molecule scattering in vacuum from
$\lambda_\varphi=\partial^2u/\partial\rho^2|\rho_0$. From
Eq.~(\ref{23}), we infer for the derivative of the potential (primes
denote derivatives with respect to $\rho$) for the simple case of
$Q_\varphi=1$ and $A_\psi=Z_\psi=Q_\psi=1$,
\begin{widetext}
\begin{eqnarray}
  \frac{\partial}{\partial k^2} {u'}\hB \hspace{-0.1cm}=
  -\frac{1}{4\pi^2 }\!\int\limits_0^\infty\! dx
  \frac{ x^2}{\sqrt{A_\varphi x^2\! +2k^2\!+ u'\!+ 2\rho u''}}
  \frac{1}{\sqrt{A_\varphi x^2\! +2k^2\!+ u'}}
  \left[\!\frac{\rho u^{'' \,2}}{A_\varphi x^2\!+2k^2\!+ u'} 
    - \frac{3\rho u^{'' \,2} + 2\rho^2 u'' u'''}{A_\varphi x^2\!
      +2k^2\!+ u'\! + 2 \rho u''}\!\right]\!\!. 
\end{eqnarray}
\end{widetext}
This contribution vanishes for $\rho =0$, and we find in the symmetric
phase no influence on the running of $m_\varphi^2$.

The bosonic contribution to the running of $\lambda_\varphi=u''(0)$,
\begin{eqnarray}\label{63}
  \frac{\partial}{\partial k^2} \lambda_\varphi\hB &=& 
  \frac{\lambda_\varphi^2}{2\pi^2}\int\limits_0^\infty
  dx \frac{x^2}{(A_\varphi x^2+2k^2 + m_\varphi^2)^2}\\
  &=& \frac{1}{8\pi}\frac{\lambda_\varphi^2}{ A_\varphi^{3/2}
    (2k^2+m^2_\varphi)^{1/2}}, \nonumber 
\end{eqnarray}
combines with the fermionic contribution and the anomalous dimension
\begin{eqnarray}
  \frac{\partial}{\partial k^2} \lambda_\varphi &=& 
  \frac{h_\varphi^2\lambda_\varphi}{32\pi} (k^2-\sigmah)^{-3/2} +
  \frac{\partial}{\partial k^2} \lambda_\varphi\hF +
  \frac{\partial}{\partial k^2} \lambda_\varphi\hB,\nonumber\\
  \frac{\partial}{\partial k^2} \lambda_\varphi\hF &=& -
  \frac{3h_\varphi^4}{8\pi^2 }\int\limits_0^\infty
  dx \frac{x^2}{(x^2+k^2 -\sigmah)^4} \\\nonumber
  &=& - \frac{3h_\varphi^4}{256 \pi } \quad (k^2-\sigmah)^{-5/2}.
\end{eqnarray}
Similar to the flow equation for $\lambda_\varphi$, the equation for
the gradient coefficient $A_\varphi$ decouples from eqs.
(\ref{A4c},\ref{mflowT0}) in the symmetric phase at $T=0$ and reads
\begin{eqnarray}
  \frac{\partial}{\partial k^2} A_\varphi  =
  \frac{h_\varphi^2}{64\pi}(k^2-\sigmah)^{-3/2}\left( A_\varphi -
    \frac{1}{2}\right). \label{eq:aphi1} 
\end{eqnarray}
There are no boson contributions in this regime. The equation is
solved by $A_\varphi = 1/2$ for our initial condition
$A_{\varphi,\text{in}}=1/2$ (the initial condition is actually
irrelevant, since $A_\varphi=1/2$ is an IR attractive fixed point of
Eq.~\eqref{eq:aphi1}).

It is interesting to compare the scattering length for molecules
$a_{\text{M}} = \bar{\lambda}_\varphi M /(2\pi)$ with the fermionic
scattering length $a$ (\ref{FermScatt}). For this purpose, we are
interested in the flow of the ratio $R_{\text{a}} = a_{\text{M}}/a =
2\lambda_\varphi/ \lambda_{\psi,\text{eff}}$. For the sake of a simple
analytic discussion, we take $A_\varphi =1/2, \lambda_{\psi,0}=0$ and
consider the broad-resonance limit. Let us first consider
\begin{eqnarray}
  \tilde{R} = \frac{8\lambda_\varphi(k^2-\sigmah)}{h_\varphi^2}.
\end{eqnarray}
which approaches $R_a$ in the infrared limit $k\to 0$, $
\tilde{R}(k\to 0) =R_a$. With $K^2=k^2 -\hat\sigma$, we find the flow
equation
\begin{eqnarray}
  K^2\frac{\partial \tilde R }{\partial K^2} &=& \tilde{R} - \frac{8
    K^4\lambda_\varphi}{h_\varphi^4}\frac{\partial h_\varphi^2}{\partial
    K^2}
  + \frac{8 K^4}{h_\varphi^2}\frac{\partial \lambda_\varphi}{\partial
    K^2}\\
  &=&\tilde{R} + \frac{1}{8\pi}\lambda_\varphi K -
  \frac{3}{32\pi}\frac{h_\varphi^2}{K}
  + \frac{2\sqrt{2}\lambda_\varphi^2 K^4}{\pi h_\varphi^2
    \sqrt{2k^2+m^2_\varphi}}\nonumber\\
  &=& \tilde{R} + \frac{h_\varphi^2}{32\pi K}\left( \frac{1}{2} \tilde{R}
    - 3 + \sqrt{\frac{k^2-\sigmah}{k^2+m^2_\varphi/2}} \tilde{R}^2
  \right).\nonumber 
\end{eqnarray}
Inserting the fixed point behaviour $h_\varphi^2 = 32 \pi K$ yields
\begin{eqnarray}\label{68}
  K^2 \frac{\partial \tilde{R}}{\partial K^2} = \frac{3}{2} (\tilde{R} -
  2)  +\sqrt{\frac{k^2-\sigmah}{k^2+m^2_\varphi/2}}\, \tilde{R}^2. 
\end{eqnarray}
If we neglect the contribution from the molecule fluctuations (the
last term in Eq.~(\ref{68})) we find an infrared stable fixed point at
$\tilde R =2$. For this fixed point, one obtains for $k\to
0,\sigmah\to\sigmah_{\text{A}}$:
\be\label{b8} \lambda_\varphi=-\frac{h^2_\varphi}{4\sigmah_{\text{A}}}
=-\frac{h^2_\varphi}{2\hat\epsilon_{\text{M}}}, \ee
to be compared with
$\lambda_{\psi,\text{eff}}=-h^2_\varphi/(2\hat\epsilon_{\text{M}})$, as
appropriate for the molecule phase. For the atom scattering in the
molecule phase, the propagator of the exchanged molecule has to be
evaluated for nonzero $\omega=-\hat{\epsilon}_{\text{M}}/2$,
cf.~Eq.~\eqref{60}. We discuss this issue in the appendix. We
conclude that this fixed point corresponds to the mean-field result
$a_{\text{M}}=2a,R_\text{a}=2$, see also \cite{Diehl:2007th}.

The bosonic fluctuations will lower the infrared value of $\tilde R$.
Using the fixed point for $h^2_\varphi$ and the expression (\ref{EE})
for the binding energy in Eq.~(\ref{ZAB}), one finds
\be\label{b8a} k^2+\frac{m^2_\varphi}{2}=3k^2-\hat\epsilon_{\text{M}}
-\sqrt{-\hat\epsilon_{\text{M}}}\sqrt{2k^2-\hat\epsilon_{\text{M}}}.
\ee
At the resonance point $\hat{\epsilon}_{\text{M}}=0$, the flow
equation for $\tilde R$ becomes
\begin{eqnarray}\label{70}
  k^2 \frac{\partial \tilde R}{\partial k^2} = -3 +\frac{3}{2} 
  \tilde{R}+ \frac{1}{\sqrt{3}} \tilde{R}^2.
\end{eqnarray}
The infrared-stable fixed point now occurs for
\begin{eqnarray}\label{CritR}
  \tilde R_* = \frac{\sqrt{3}}{2}\Big( \sqrt{ \frac{9}{4} 
    + \frac{12}{\sqrt{3}}} - \frac{3}{2}\Big) \approx 1.325.
\end{eqnarray}
Precisely on the Feshbach resonance ($B=B_0$), the asymptotic behaviour
obeys the scaling form
\begin{eqnarray}\label{LambdaScaling}
  \lambda_\varphi =\frac{h^2_\varphi\tilde R_*}{8k^2}
  = \frac{4\pi\tilde{R}_*}{k}.
\end{eqnarray}

For $B<B_0$, the mass term $m_\varphi^2$ is smaller as compared to the
critical value for $B=B_0$. For a given $K$ the expression
$\sqrt{2k^2+m^2_\varphi}=\sqrt{2K^2+2\sigmah+m^2_\varphi}$ is
therefore smaller and the term $\sim\tilde R^2$ in the flow Eq.~is
enhanced. As a consequence, $\lambda_\varphi$ turns out smaller than
the critical value (\ref{LambdaScaling}).

It is instructive to study the flow in the range where
$k^2\ll-\hat\epsilon_{\text{M}}/2,K\approx
\sqrt{-\hat\epsilon_{\text{M}}/2}+k^2/\sqrt{-2\hat\epsilon_{\text{M}}}$
and therefore
\be\label{c8a} k^2+\frac{m^2_\varphi}{2}\approx 2k^2.  \ee
The flow equation for $\tilde R$ reads
\be\label{c8b} k^2\frac{\partial}{\partial k^2}\tilde R
=\left\{\frac{3}{2}(\tilde R-2)
  +\frac{\sqrt{2k^2-\hat\epsilon_{\text{M}}}}{2k}\tilde R^2\right\}
\frac{k^2}{k^2-\hat\epsilon_{\text{M}}/2}.  \ee For $k\to0$, the
second term dominates, \be\label{BDR} \frac{\partial\tilde R}{\partial
  k}=\frac{\sqrt{2}\tilde R^2}{\sqrt{k^2-\hat\epsilon_{\text{M}}/2}},
\ee and the running stops for $k\to 0$, resulting in \be\label{c8c}
\tilde R^{-1}(k=0)\approx \tilde R^{-1}(k_{\text{tr}})+\sqrt{2}
\text{Arsinh}\left(\frac{\sqrt{2}k_{\text{tr}}}{
    \sqrt{-\hat\epsilon_{\text{M}}}} \right).  \ee
Here, $k_{\text{tr}}$ is a typical value for the transition from the
scaling behaviour (\ref{70}) for $k^2\gg-\hat\epsilon_{\text{M}}/2$ to
the boson dominated running (\ref{BDR}) for
$k^2\ll-\hat\epsilon_{\text{M}}/2$. We may give a rough estimate using
$k_{\text{tr}}=\sqrt{-\hat\epsilon_{\text{M}}/2}$ and $\tilde
R(k_{\text{tr}})=\tilde R_*,$
\be\label{c8d} \tilde R(0)=\frac{\tilde R_*}{1+\sqrt{2}
  \text{Arsinh}(1) \tilde R_*}\approx0.5.  \ee

The value for a numerical solution turns out somewhat higher and we
find for $k\to0$
\be
\label{c8e} 
\tilde R( k\to 0)= \frac{a_\text{M}}{a} =  0.81. 
\ee
This is mainly due to the presence of fermion fluctuations which
enhance $\tilde R$. 

Following the flow of $\lambda_\varphi$ in a simple
truncation thus predicts $a_{\text{M}}/a\simeq 0.81$, in qualitative
agreement with rather refined quantum mechanical computations
\cite{Petrov04}, numerical simulations \cite{Giorgini04}, and sophisticated diagrammatic approaches \cite{Kagan05}, $a_{\text{M}}/a\simeq 0.6$. The resummation obtained from our flow equation is equivalent to the one performed in \cite{AAAAStrinati}, yielding $a_{\text{M}}/a\simeq 0.78\pm4$. Functional RG flows in the present
truncation with optimised cutoffs \cite{Litim:2000ci,Pawlowski:2005xe}
lead to $a_{\text{M}}/a\simeq 0.71$ \cite{Diehl:2007th}. It should be
possible to further improve the quantitative accuracy by extending the
truncation, for example by including an interaction of the type
$\psi^\dagger\psi\varphi^*\varphi$ corresponding to atom-molecule scattering. Already at this stage, it is clear
that the final value of $R_{\text{a}}$ results from an interesting interplay
between the fermionic and bosonic fluctuations. In a model describing
only interacting bosonic particles (atoms or molecules), the flow of
$\lambda_\varphi$ can be taken from Eq.~(\ref{63}) for
$m^2_\varphi=0,A_\varphi=\frac12$
\be\label{BBB}
\frac{\partial\lambda_\varphi}{\partial
  k}=\frac{1}{2\pi}\lambda^2_\varphi.
\ee
(For an application to bosonic atoms, one should recall that our
normalisation of the particles corresponds to atom number 2 and mass
$2M$.) The bosonic fluctuation effects have the tendency to drive
$\lambda_\varphi$ towards zero, according to the solution of
Eq.~(\ref{BBB}) for $k\to 0$,
\be\label{c8f} \lambda_\varphi= \frac{\lambda_{\varphi,\text{in}}}{1+
  \lambda_{\varphi,\text{in}}k_{\text{in}}/2\pi}.  \ee
This solution reflects the infrared-stable fixed point for
$\tilde\lambda_\varphi=\lambda_\varphi k$,
\be\label{c8g}
(\tilde\lambda_\varphi)_*=0,
\ee
resulting from
\be\label{c8h} k\frac{\partial}{\partial k}\tilde\lambda_\varphi=
\tilde\lambda_\varphi+\frac{\tilde\lambda^2_\varphi}{2\pi}.  \ee

On the other hand, the fermion fluctuations have the opposite
tendency: they generate a non-vanishing $\lambda_\varphi$ even for an
initial value $\lambda_{\varphi,\text{in}}=0$ and slow down the
decrease of $\lambda_\varphi$ as compared to Eq.~(\ref{BBB}) in the
range of small $k$. The resulting $R_{\text{a}}$ results from a type of balance
between the opposite driving forces. If the scaling behaviour
(\ref{70}) is approximately reached during the flow, the final value
of $R_{\text{a}}$ is a universal ratio which does not depend on the microscopic
details. This is a direct consequence of the ``loss of microscopic
memory'' for the fixed point (\ref{CritR}). 

\section{Universality}
\label{universality}

The essential ingredient for the universality of the BEC-BCS crossover
for a broad Feshbach resonance is the fixed point in the
renormalisation flow for large $\tilde h^2_\varphi$ (fixed point $(A)$
in sect. \ref{density}). This fixed point is infrared stable for all
couplings except one relevant parameter corresponding to the detuning
$B-B_0$. For $T=0$, this fixed point always dominates the flow at
$B=B_0$ as long as effects from a nonzero density can be
neglected. However, for any $k_{\text{F}}\neq 0$ the flow will finally
deviate from the scaling form due to the occurrence of a condensate
$\rho_0\neq0$. Typically, this happens once $k$ reaches the gap for
single fermionic atoms, $k\approx\Delta=h_\varphi\sqrt{\rho_0}$. (If
we use $\hat k=k_{\text{F}}$ the relevant scale is
$k=16\Omega_c/(3\pi).)$ For $k\ll\Delta$, the contribution from
fermionic fluctuations to the flow gets suppressed. Also the
contribution from the ``radial mode'' of the bosonic fluctuations will
be suppressed due to a mass-like term $m^2_\varphi\approx
2\lambda_\varphi\rho_0$. Only the Goldstone mode corresponding to the
phase of $\varphi$ will have un-suppressed fluctuations.

For nonzero density it is convenient to express all quantities in
units of the Fermi momentum $k_{\text{F}}$ and the Fermi energy
$\epsilon_{\text{F}}$, i.e., to set $\hat k=k_{\text{F}}$. As compared
to the scaling form at zero density the flow equations are now
modified by two effects. They concern the influence of $\rho_0\neq0$
and the nonzero value for $\sigmah=\sigmah_*$ which corresponds to
$B=B_0$. (Typically, this value $\sigmah_*$ is positive, as opposed to
$\sigmah_*=0$ for the zero density case.) Nevertheless, all these
modifications only concern the flow for small $k$. If the flow for
large $k$ has already approached the fixed point close enough no
memory is left from the microscopic details. We can immediately
conclude that all physical quantities are universal for $B=B_0$.

The same type of arguments applies for nonzero temperature. For $\hat
T\neq0$, a new effective infrared cutoff is introduced for the
fermionic fluctuations. Also the contributions from the bosonic
fluctuations get modified. Again, this only concerns the flow for
$k\ll\pi\hat T$. The loss of microscopic memory due to the attraction
of the flow for large $k$ towards the fixed point remains effective.

In a similar way, we can consider the flow for $B\neq B_0$. Away from
the exact location of the resonance, the observable quantities will
now depend on the relevant parameter. The latter can be identified
with the concentration $c=ak_{\text{F}}$ \cite{Diehl:2005an}. Still,
the deviation from the scaling flow only concerns small $k$, namely
the range when the first term in Eq.~(\ref{ZAB}) becomes
important. This happens for
\be\label{c8i}
k\approx\frac{16\pi M\bar\mu|B-B_0|}{\hat h^2_{\varphi,0}k^2_F}=\frac1c.
\ee
If the flow for large $k$ has been attracted close enough to the
fixed point all physical quantities at nonzero density are universal
functions of two parameters, namely $c$ and $\hat
T=T/\epsilon_{\text{F}}$. These functions describe all broad resonances.

\section{Deviations from universality}
\label{deviations}

The functional flow equations also allow for simple quantitative
estimates for the deviations from universality for a given atom gas.
Typically, these corrections are power suppressed
$\sim(\Lambda_{\text{IR}}/\Lambda_{\text{UV}})^p$. The microscopic
scale $\Lambda_{\text{UV}}$ roughly denotes the inverse of the
characteristic range of van der Waals forces and we may associate
$\Lambda_{\text{UV}}\simeq 1/(100 a_{\text{A}})$ with $a_{\text{A}}$
the ``radius'' of the atoms. More precisely, $\Lambda_{\text{UV}}$
corresponds to the scale at which the attraction to the fixed point
sets in. In some cases, this may be substantially below the van der
Waals scale, as for the case of ${}\kal$, as we will argue below. The
relevant infrared scale $\Lambda_{\text{IR}}$ depends on the density,
$\Lambda_{\text{IR}}=L(c,\hat T)k_{\text{F}}$. In view of the
discussion in the preceding section, we approximate $L$ by the highest
value of $k$ where the deviation from the scaling form of the flow
equations sets in, i.e.,
\be
\label{c8j}
L=\max\{|c^{-1}|,\pi\hat T,\sqrt{|\sigmah|},\hat
\Delta\}.
\ee
We note that $\sigmah$ and
$\hat\Delta=h_\varphi\sqrt{\rho_0}$ depend on $c$ and
$\hat T$.

The power $p$ of the suppression factor reflects the ``strength of
attraction'' of the fixed point. More precisely, we may consider the
flow of various dimensionless couplings $\alpha_i$. If we denote their
values at the fixed point by $\alpha_{i*}$, the flow of small
deviations $\delta\alpha_i=\alpha_i-\alpha_{i_*}$ is governed by the
``stability matrix'' $A_{ij}$
\be\label{c8k}
k\frac{\partial}{\partial k}\delta\alpha_i=A_{ij}\delta\alpha_j.
\ee
The relevant parameter corresponds to the negative eigenvalue of
$A$. All other eigenvalues of $A$ are positive, and $p$ is given by
the smallest positive eigenvalue of $A$.

Typical couplings $\alpha_i$ are $(\tilde\lambda_\psi,\tilde
h^2_\varphi,\tilde m^2_\varphi,\tilde R)$. For these couplings, the
stability matrices reads for fixed point (A):
\be\label{c8l}
A=\left(\begin{array}{llll}
1,&0,&0,&0\\
8,&1,&0,&0\\
0,&\frac{1}{4\pi},&-1,&0\\
0,&e_1,&e_2,&e_3
\end{array}\right)
\ee
with $e_1=(\tilde R_*-6+2\tilde R^2_*)/(32\pi),e_2=-\tilde
R^2_*/(6\sqrt{3}), e_3=3+4\tilde R_*/\sqrt{3}$.  The eigenvalues are
$(1,1-1,e_3)$ and we conclude $p=1$. This situation is not expected to
change if more couplings like $A_\varphi,A_\psi$ or the $\rho^3$ term
in the potential are included.

We may use our findings for a rough estimate for the range in $c^{-1}$
and $\hat T$ for which deviations from universality are smaller than
$1\%$. From our previous considerations this holds for
\be\label{c8m}
|c^{-1}|\lesssim\frac{\Lambda_{\text{UV}}}{100k_{\text{F}}}~,\qquad 
~\sqrt{\hat T}\lesssim
\frac{\Lambda_{\text{UV}}}{300k_{\text{F}}}.
\ee
As a condition for $\Lambda_{\text{UV}}$, we require that all couplings
are in a wider sense in the ``vicinity'' of the fixed point at the
scale $k_{\text{UV}}=\Lambda_{\text{UV}}/k_{\text{F}}$. Of course, one
necessary condition is $\Lambda_{\text{UV}}<\Lambda$. Also, the flow
equations must be meaningful for $k<k_{\text{UV}}$. While this poses
no restriction on Li where $\lambda_{\psi,0}<0$, the case of $K$
with positive $\lambda_{\psi,0}$ requires $k_{\text{UV}}\hat c_0<1$ or
$k_{\text{UV}}\lesssim 20$. If we would try to extrapolate the flow to
even higher $k$ the value of $\hat h^2_\varphi(k)$ would diverge as
can be seen from Eq.~(\ref{A4f}).  As a further condition for the flow
to be governed by the universal behaviour of the fixed point we require
$Z_\varphi -1\gtrsim 1$. This tells us that the notion of a di-atom
state $\varphi$ starts to be independent of the detailed microscopic
properties. For Li and K, this happens for scales only somewhat below
$k_{\text{in}}$. 

A reasonable estimate at this stage may be
$k^{\text{(Li)}}_{\text{UV}}=800, k^{\text{(K)}}_{\text{UV}}=15$. One
percent agreement with the universal behaviour would then hold for
\be\label{c8n}
\begin{array}{lll}
\text{Li}:&|c^{-1}|\lesssim 8~,\qquad  &\sqrt{\hat T}\lesssim 3,\\
\text{K}:&|c^{-1}|\lesssim 0.15~,&\sqrt{\hat T}\lesssim 0.05,
\end{array}
\ee
which corresponds to
\be
\begin{array}{lll}
\text{Li}:&|a k_\text{F}|\gtrsim 0.13~,\qquad  & T/T_\text{F} \lesssim 9,\\
\text{K}:&|a k_\text{F}|\gtrsim 6.7~,&T/T_\text{F} \lesssim 0.0025.
\end{array}
\ee
Already at this stage, we conclude that the universal behaviour for K
covers only a much smaller range in $c$ and $\hat T$ as compared to
Li. In particular, experiments with Li are indeed performed within the
universal regime which extends far off resonance and to temperatures
well above the quantum degenerate regime $T/T_\text{F} \approx 1$,
while for $K$ it will be hard to reach the truly universal domain,
since the lowest available temperatures range down to $T/T_\text{F}
\approx 0.04$, and the magnetic field tuning is too low to resolve a
regime where $|a k_\text{F}|\gtrsim 6.7$.

At first sight, the limitation to a rather small value
$k_{\text{UV}}^{\text{(K)}}=15$ for $\kal$ may seem to be of a
technical nature, enforced by the breakdown of the flow for
$\lambda_\psi$ for $k>k_{\text{UV}}^{\text{(K)}}$. However, this
technical shortcoming most likely reveals the existence of some
additional physical scale close to $k_{\text{UV}}^{\text{(K)}}$, as,
for example, associated to a further nearby resonance state not
accounted for in our simple model. Indeed, if our model (without
modifications and additional effective degrees of freedom) would be
valid for $k\gg k_{\text{UV}}^{\text{(K)}}$, one could infer an
effective upper bound for $\tilde\lambda_\psi$ from its flow. Starting
at some $k_{\text{in}}$ with an arbitrarily large positive
$\tilde{\lambda}_{\psi,\text{in}}$, the value of
$\tilde{\lambda}_\psi(k)$ would be renormalised to a finite value,
bounded by $\tilde{\lambda}_\psi(K=0)<8\pi/K_{\text{in}}$
(cf. Eq.~\eqref{A4d}). This would lead to a contradiction  with the
observed value unless $K_{\text{in}}$ is sufficiently low. The
observed value of $\tilde{\lambda}_{\psi,0}$ therefore implies the
existence of a scale near $k_{\text{UV}}^{\text{(K)}}$ where our
simple description needs to be extended. This issue is very similar
to the ``triviality bound'' in the standard model for the electroweak
interactions in particle physics.

As further concern, one may ask if $\tilde h^2_\varphi(k_{\text{UV}})$
is already close enough to the fixed point value $32\pi$ and if
$\tilde\lambda_\psi(k_{\text{UV}})$ is close enough to zero. This is
an issue, since we know of the existence of a different fixed point
for narrow Feshbach resonances at $\tilde
h^2_\varphi=\tilde\lambda_\psi=0$. At the scale $k_{\text{UV}}$, the
flow should not be in the vicinity of this ``narrow-resonance fixed
point'' anymore, but at least in the crossover region towards the
``broad-resonance fixed point''. If we evaluate the couplings at the
scales
$k^{\text{(Li)}}_{\text{UV}}=800~,~k^{\text{(K)}}_{\text{UV}}=15$ we
obtain for $c^{-1} = 0$
\begin{eqnarray}
\text{Li}&:&\tilde h^2_\varphi (k_{\text{UV}})= 5.00 \cdot 10^{-3},
\quad\tilde\lambda_\psi(k_{\text{UV}})= -25.0, \nonumber \\
&& \quad\tilde m^2_\varphi(k_{\text{UV}})=
6.07  \cdot 10^{-2};\nonumber \\
\text{K}&:&\tilde h^2_\varphi(k_{\text{UV}})= 4.53 \cdot 10^{3},
\quad\tilde\lambda_\psi(k_{\text{UV}})=58.8, \nonumber\\
&&\quad\tilde m^2_\varphi(k_{\text{UV}})= 54.0. 
\label{c8o}
\end{eqnarray}
For K, the relevance of the broad-resonance fixed point seems
plausible, but the small value of $\tilde h^2_\varphi(k_{\text{UV}})$
for Li may shed doubts. However, one should keep in mind that for a
strong enough $\lambda_{\psi,\text{eff}}$ the distribution between
$\lambda_\psi$ and $-h^2_\varphi/m^2_\varphi$ concerns mainly the
dependence of the scattering length on the magnetic field $B$ and not
so much the physics for a given $a$ or $c$. Indeed, we could absorb
$\lambda_\psi$ by partial bosonisation (Hubbard-Stratonovich
transformation) in favour of a change of $h^2_\varphi$ and
$m^2_\varphi$. Up to terms $\sim q^4$, which are sub-leading for the
low-momentum physics, a model with given
$\lambda_\psi,h^2_\varphi,m^2_\varphi$ is equivalent to a model with
$\lambda'_\psi=0$, but new values $h^{'2}_\varphi$ and
$m'_\varphi{}^2$ related to the original parameters by
\ba\label{c8p}
h^{'2}_\varphi&=&h^2_\varphi-2\lambda_\psi m^2_\varphi+\frac{\lambda^2_\psi m^4_\varphi}{h^2_\varphi}\nonumber\\
m^{'2}_\varphi&=&m^2_\varphi-\frac{\lambda_\psi m^4_\varphi}{h^2_\varphi}.
\ea
The value of the new Yukawa coupling for $\text{Li},\tilde
h^{'2}_\varphi(k_{\text{UV}})= 4.65 \cdot 10^2$, is much larger and seems
acceptably close to the broad resonance fixed point. Using partial
bosonisation and the technique of rebosonisation during the flow
\cite{Gies01}, it may actually be possible to treat $\lambda_\psi$ as
a redundant parameter, thus enlarging the ``space of attraction'' of
the fixed point (A). In this context, we note that we should include
the contribution of the bosonic fluctuations to the running of
$\lambda_\psi$ for our truncation. This will shift the precise
location of the fixed points. In a language with rebosonisation where
$\tilde\lambda_\psi$ remains zero, these additional contributions will
be shifted into the flow of $\tilde m^2_\varphi$ and $\tilde
h^2_\varphi$.

Our estimate for the deviations from universality concerns only the
overall suppression factor. Detailed proportionality coefficients
$c_\mathcal{O}$ for the deviation of a given observable $\mathcal O$
from the universal strong-resonance limit, $\delta
\mathcal{O}=c_\mathcal{O} \Lambda_{\text{IR}}/\Lambda_{\text{UV}}$,
depend on the particular observable. The deviations from universality
should therefore be checked by explicit solutions of the flow equation
for Li and K.

\section{Conclusions}
\label{conclusion}

In this paper, we have performed a functional renormalisation group
study for ultracold gases of fermionic atoms. We have concentrated on
four parameters: the Yukawa coupling of the molecules to atoms
$h_\varphi$, the offset between molecular and open channel energy
levels $m^2_\varphi$ which is related to the detuning, the background
atom-atom-interaction $\lambda_\psi$ and the
molecule-molecule interaction $\lambda_\varphi$. This system exhibits
a fixed point for the rescaled couplings which is infrared stable
except for one relevant parameter. This parameter can be associated
with the inverse concentration $c^{-1}=(ak_{\text{F}})^{-1}$.

Whenever for a given system the trajectories of the flow in
coupling-constant space approach this fixed point close enough, the
macroscopic physics looses the memory of the microphysical details
except for one parameter, namely $c^{-1}$. In consequence, for a
certain range in $c^{-1}$ around $0$ and for a certain range in
temperature $\hat T$ all dimensionless macroscopic quantities can only
depend on $c$ and $\hat T$. Here dimensionless quantities are
typically obtained by multiplication with appropriate powers of
$k_{\text{F}}$ or $\epsilon_{\text{F}}$.

The macroscopic quantities include all thermodynamic variables and,
more generally, all quantities that can be expressed in terms of
$n$-point functions for renormalised fields at low momenta. In
particular, the correlation functions for atoms and ``molecules'' as
well as their interactions depend only on $c$ and $\hat T$,
independently of the concrete microscopic realization of a broad
Feshbach resonance. Here ``molecules'' refers to bosonic quasi
particles as collective di-atom states which may be quite different
from the notion of microscopic or ``bare'' molecules. In a certain
sense this situation has an analogue in the universal critical
behaviour near a second order phase transition. In contrast to this,
however, the universal description includes now a temperature range
between $\hat T=0$ and substantially above the critical temperature,
and the universal quantities depend on an additional relevant
parameter $c^{-1}$. We may also compare to a quantum phase transition
at $T=0$ which is realized in function of a parameter analogous to
$c^{-1}$. In our case, universality is extended to nonzero temperature
as well.

The ``broad resonance fixed point'' is not the only fixed point of the
system. Another ``narrow resonance fixed point'' at $\tilde
h^2_\varphi=0,~\tilde\lambda_\varphi=0$ allows for an exact solution
of the crossover problem for narrow Feshbach resonances
\cite{Diehl:2005an,Radzihovsky07} and a perturbative expansion around this
solution. The narrow-resonance fixed point has two relevant parameters
$c^{-1}$ and $\tilde h^2_\varphi$. A typical flow away from this fixed
point at small enough $c^{-1}$ and $\lambda_\psi$ describes a
crossover towards the broad resonance fixed point.

For a given physical system characterised by some microscopic
Hamiltonian, it is important to determine how close it is to the
universal behaviour of the broad-resonance limit. We have performed
here a first estimate for the range in $c^{-1}$ and $\hat T$ for which
universality holds within one percent accuracy for the experimentally
studied Feshbach resonances in $\lit$ and $\kal$. It will be an
experimental challenge to verify of falsify these predictions of
universality.

\appendix*

\section{Scattering length and Cross section}

In this appendix, we collect useful formulae for the scattering of
molecules in vacuum. In quantum field theory, the scattering cross
section for identical nonrelativistic bosons is given by
\begin{eqnarray}
\sigma_{\text{B}} = \frac{1}{2}\int d\Omega \frac{d\sigma}{d\Omega}.
\end{eqnarray}
The factor $1/2$ is a convention, motivated by integration over half
the spatial angle in order to avoid double counting of the
identical particles. The differential cross section is related to
the scattering amplitude by
\begin{eqnarray}
\frac{d\sigma}{d\Omega} = \frac{M_B^2|A_{\text{B}}|^2}{16\pi^2},
\end{eqnarray}
such that, in case of isotropic (e.g. s-wave) scattering
\begin{eqnarray}
\sigma_{\text{B}} = \frac{M_B^2|A_{\text{B}}|^2}{8\pi}.\label{A3}
\end{eqnarray}
Next, we relate the above results to nonrelativistic quantum
mechanics: The scattering length $a$ is a quantity which is directly
meaningful in nonrelativistic quantum mechanics only. It measures the
strength of the $1/r$ decay of the scattered wave function at low
energies as $a/r$. Equivalently, $a$ is defined by the $l=0$ phase
shift for the scattered wave function. This definition leads to the
relation between scattering length and cross section (identical
bosons)
\begin{eqnarray}
\sigma_{\text{B}}  = 8\pi a_{\text{B}}^2.\label{A4}
\end{eqnarray}
We can use Eqs.~\eqref{A3}, \eqref{A4} in order to relate the
scattering amplitude to the scattering length or rather to define this
relation,
\begin{eqnarray}
|A_{\text{B}}| = \frac{8 \pi a_{\text{B}}}{M_B}=\frac{2
 \bar{\lambda}_\varphi}{Z_\varphi^2}. 
\end{eqnarray}
The scattering amplitude is given by the effective four-boson vertex,
as obtained by functional derivatives of the effective action,
$Z_\varphi^2 A_{\text{B}}=\Gamma^{(4)}=2\bar\lambda_\varphi$. In the limit of a
point-like interaction, $\bar{\lambda}_\varphi$ is a constant. (Note
that the notion of scattering amplitude is used to describe different
quantities in quantum mechanics and QFT.)

For fermions, the definitions are similar,
\begin{eqnarray}
\sigma_{\text{F}} = \int d\Omega \frac{d\sigma}{d\Omega}.
\end{eqnarray}
Now, the integration covers the full space angle, since the
fermions can be distinguished by their spin.  The differential cross
section is related to the scattering amplitude by
\begin{eqnarray}
\frac{d\sigma}{d\Omega} = \frac{M_{\text{F}}^2|A_{\text{F}}|^2}{16\pi^2},
\end{eqnarray}
such that for isotropic scattering
\begin{eqnarray}
\sigma_{\text{F}} = \frac{M_{\text{F}}^2|A_{\text{F}}|^2}{4\pi}.
\end{eqnarray}
For distinguishable fermions, one has in quantum mechanics
\begin{eqnarray}
\sigma  = 4\pi a_{\text{F}}^2,
\end{eqnarray}
such that the scattering amplitude and scattering length are related
by
\begin{eqnarray}
|A_{\text{F}}| = \frac{4 \pi
 a_{\text{F}}}{M_{\text{F}}}=\bar{\lambda}_{\psi,\text{eff}}
(\omega,\vec{q}).  
\end{eqnarray}

The scattering amplitude $A_\text{F}$ is given by the tree level graph
for the molecule exchange fermions plus a contribution from the
fermionic 1PI vertex $\bar\lambda_\psi$. The energy- and momentum-dependent
resonant four-fermion vertex generated by the molecule exchange reads
\begin{eqnarray}
&&\bar \lambda_{\psi,\text{eff}} (\omega,\vec q) -\bar\lambda_{\psi,0} =
- \frac{\bar h_\varphi^2(\sigma_{\text{A}})}{\bar{P}_\varphi(\omega,\vec q)} \\\nonumber
&=&- \frac{\bar h_\varphi^2(\sigma_{\text{A}})}{-\omega + \frac{\vec q^2}{4M} +
 \bar{m}_\varphi^2 + \big(\Delta \bar P_\varphi (\omega,\vec q) -
 \Delta \bar P_\varphi (0,\vec 0)\big) }\\\nonumber 
\end{eqnarray}
with
\begin{eqnarray}
\Delta \bar P_\varphi (\omega,\vec q)
=\frac{\bar{h}_\varphi^2(\sigma_{\text{A}})M^{3/2}}{4\pi}\sqrt{-\omega
  -2 \sigma_{\text{A}}+ \frac{\vec q^2}{4M}}. 
\end{eqnarray}

With $\vec{q}_1,\vec{q}_2$ the momenta of the scattered atoms, one
finds for the momentum and energy of the exchanged molecule
$\vec{q}=\vec{q}_1+ \vec{q}_2$,
$\omega=\vec{q}_1^2/(2M)+\vec{q}_2^2/(2M) -2\sigma_{\text{A}}$. Here,
we take into account the binding energy -- the energy of an incoming
atom is $-\sigma_{\text{A}} + \vec{q}_i^2/(2M)$. We now consider the
limit $\vec{q}_i\to 0$ and work in the broad-resonance limit
$\bar{h}_\varphi^2\to\infty$. The vacuum in the molecule phase is
characterised by $\bar{m}_\varphi^2=0$, and we end up with
\begin{equation}
\bar{\lambda}_{\psi,\text{eff}} -\bar{\lambda}_{\psi,0}
=\frac{4\pi}{M^{3/2} \sqrt{-2\sigma_{\text{A}}}}. 
\end{equation}
One infers for the scattering length of atoms in the molecule phase
\begin{eqnarray}
a = \frac{M
  \bar{\lambda}_{\psi,\text{eff}}(-2\sigma_{\text{A}},\vec{0})}{4\pi}
=   \frac{1}{\sqrt{ -2 M\sigma_{\text{A}}}} + a_0.
\end{eqnarray}

The relation between our rescaled quantities and the scattering
amplitudes is given by
\begin{eqnarray}
\hat{\lambda}_\varphi &=& 2M k_{\text{F}} \bar{\lambda}_\varphi, \quad
\lambda_\varphi=\frac{\hat{\lambda}}{Z_\varphi^2}, \quad
\hat{\lambda}_{\psi,\text{eff}} =2M k_{\text{F}}
\bar{\lambda}_{\psi,\text{eff}}, \nonumber\\
\hat{h}_{\varphi,0}^2 &=& \frac{4M^2}{k_{\text{F}}}\,
\bar{h}_\varphi^2, \quad
\hat{\sigma}_{\text{A}}=
\frac{\sigma_{\text{A}}}{\epsilon_{\text{F}}}, \quad \hat{m}_\varphi^2
= \frac{\bar{m}_\varphi^2}{\epsilon_{\text{F}}}. \label{someequation}
\end{eqnarray}
With $M_{\text{B}}=2M$, the ratio between molecular and atomic
scattering length ($a_{\text{M}}=a_{\text{B}}$, $a_{\text{M}}
k_{\text{F}}=\lambda_\varphi/(4\pi)$) therefore reads
\begin{equation}
\frac{a_{\text{M}}}{a} =
\frac{2\bar{\lambda}_\varphi/Z_\varphi^2}{\bar{\lambda}_{\psi,\text{eff}}
  (\omega,\vec{q})} 
=\frac{2{\lambda}_\varphi}{{\lambda}_{\psi,\text{eff}}
  (-2\sigma_{\text{A}},\vec{q})}. \label{someotherequation}
\end{equation}
Using the definition of the renormalised Yukawa coupling $h_\varphi^2
=\hat{h}_\varphi^2/Z_\varphi$, we may express
\begin{equation}
\lambda_{\psi,\text{eff}}(-2\sigma_{\text{A}},0)-\lambda_{\psi,0} =
\frac{8\pi}{\sqrt{-\hat{\sigma}_{\text{A}}}} 
=-\frac{h_\varphi^2}{4\hat{\sigma}_{\text{A}}} 
=-\frac{h_\varphi^2}{2\hat{\epsilon}_{\text{M}}}.
\label{yetanotherequation}
\end{equation}
In the last two expressions, we have used the leading terms in
Eqs.~\eqref{A4f}, \eqref{45} for $K=
\sqrt{-\hat{\sigma}_{\text{A}}}\to 0$.

\begin{acknowledgments}
We thank S.~Fl\"orchinger, H.C.~Krahl, M.~Scherer and P.~Strack for
useful discussions. H.G.~acknowledges support by the DFG under
contract Gi 328/1-3 (Emmy-Noether program).
\end{acknowledgments}

\bibliographystyle{apsrev}
\bibliography{Citations}

\end{document}